\documentclass[twocolumn,floats,floatfix,
    showpacs,prd,superscriptaddress,
    nofootinbib]{revtex4-2}

\usepackage{graphicx,epsfig}
\usepackage{amssymb,amsmath,amsthm,amsfonts,mathtools}
\usepackage{bm}
\usepackage[inline]{enumitem}
\usepackage{tensor}
\usepackage[usenames,dvipsnames]{xcolor}
\usepackage{url}
\usepackage{multirow}
\usepackage{array}
\usepackage{xspace}
\usepackage{comment}
\usepackage[caption=false]{subfig}
\usepackage{booktabs}
\usepackage{soul}
\usepackage{mathrsfs}

\usepackage{float}

\usepackage{accents}
\usepackage[linktocpage,breaklinks]{hyperref}
\hypersetup{colorlinks=true,
            citecolor=NavyBlue,
            linkcolor=NavyBlue,
            urlcolor=NavyBlue}

\definecolor{lightred}{RGB}{255, 120, 120}

\def\newacronym#1#2#3{\gdef#1{\gdef#1{#2\xspace}#3 (#2)\xspace}}
\def\bh#1{black hole#1 (BH#1)\gdef\bh{BH}}
\def\GW#1{gravitational wave#1 (GW#1)\gdef\GW{GW}}
\newacronym{\qbs}{QBS}{quasi-bound state}
\def\qnm#1{quasi-normal mode#1 (QNM#1)\gdef\qnm{QNM}}
\newacronym{\vsh}{VSH}{vector spherical harmonics}
\newacronym{\lfkks}{LFKKS}{Lunin-Frolov-Krtou\v{s}-Kubiz\v{n}\'{a}k-Santos}
\def\ode#1{ordinary differential equation#1 (ODE#1)\gdef\ode{ODE}}
\def\pde#1{partial differential equation#1 (PDE#1)\gdef\pde{PDE}}

\def\dif{\textrm{d}}

\def\pc{\text{pc}}
\def\ev{\text{eV}}
\def\Gev{\text{GeV}}
\def\yr{\text{yr}}
\def\msun{M_\odot}
\def\cm{\text{cm}}

\def\Lie{\mathcal{L}}

\def\re{\text{Re}\,}

\begin{document}

\title{Kerr black holes with vector dark matter hair}

\author{Fredric~Hancock}
\email{foh3@illinois.edu}
\affiliation{The Grainger College of Engineering,
Department of Physics \& Illinois Center for Advanced Studies of the Universe, University of Illinois Urbana-Champaign, Urbana, Illinois 61801, USA}

\author{Helvi Witek}\email{hwitek@illinois.edu}
\affiliation{The Grainger College of Engineering,
Department of Physics \& Illinois Center for Advanced Studies of the Universe, University of Illinois Urbana-Champaign, Urbana, Illinois 61801, USA}
\affiliation{Center for AstroPhysical Surveys, National Center for Supercomputing Applications, University of Illinois Urbana-Champaign, Urbana, IL, 61801, USA}

\begin{abstract}
    We consider a massive vector-field perturbation on a Kerr black hole background. We fix the field's density at the edge of the black hole sphere of influence to model a bath of wave cold dark matter, from which the black hole can accrete endlessly. We employ the approach of Lunin, Frolov, Krtou\v{s}, Kubiz\v{n}\'{a}k, and Santos to separate the equations, and solve them with a mix of analytical and numerical techniques. We find that the field forms a density spike around the black hole with profile $\rho \sim r^{-3/2}$, consistent with dark matter spikes studied in the literature, which is not significantly distorted by spacetime rotation, except near the black hole's ergoregion. For any nonzero spin, an additional superradiant regime appears, in which co-rotating modes extract mass and angular momentum from the black hole. We find a mass \textit{extraction} rate as high as $10\, M_\odot/\yr$ in the superradiant regime for a $10^9\, \msun$ black hole. For large field masses, the rate at which the black-hole mass grows behaves similarly to the Schwarzschild case, reaching $\sim 100\, M_\odot/\yr$ for a $10^9\, \msun$ black hole. We compare these results to their scalar-field counterparts and discuss how they mix with the superradiant instability.
\end{abstract}

\maketitle
\tableofcontents


\section{Introduction}
\label{sec:intro}

The LIGO-Virgo-KAGRA \GW{} observatories have brought us into an unprecedented era of gravitational physics, in which mergers of \bh{s} and neutron stars are detected on a weekly basis~\cite{LIGOScientific:2025snk, LIGOScientific:2026jgl}
At current detector sensitivities, however, these observations are largely compatible with vacuum environments---i.e., two compact objects surrounded by no matter~\cite{CanevaSantoro:2023aol}. 
This may change with the next generation of \GW{} detectors, particularly with LISA, expected to launch in the mid 2030s~\cite{LISA:2017pwj}. 
LISA will be able to observe \bh{} binaries of total mass $10^4\, \msun$--$10^8\, \msun$ through numerous orbits of their inspiral, with enough sensitivity to potentially reveal 
a phase shift of the \GW{} signal, due to interactions with matter in their environment,
especially for extreme mass ratio binaries~\cite{Barausse:2020rsu, LISA:2022kgy, Duque:2023seg, Gliorio:2025cbh}. 
Before LISA launches, then, it is imperative that we understand precisely how the various potential components of a \bh{'s} environment would affect this signal. 

\begin{figure}
    \centering
    \includegraphics[width=0.9\linewidth]{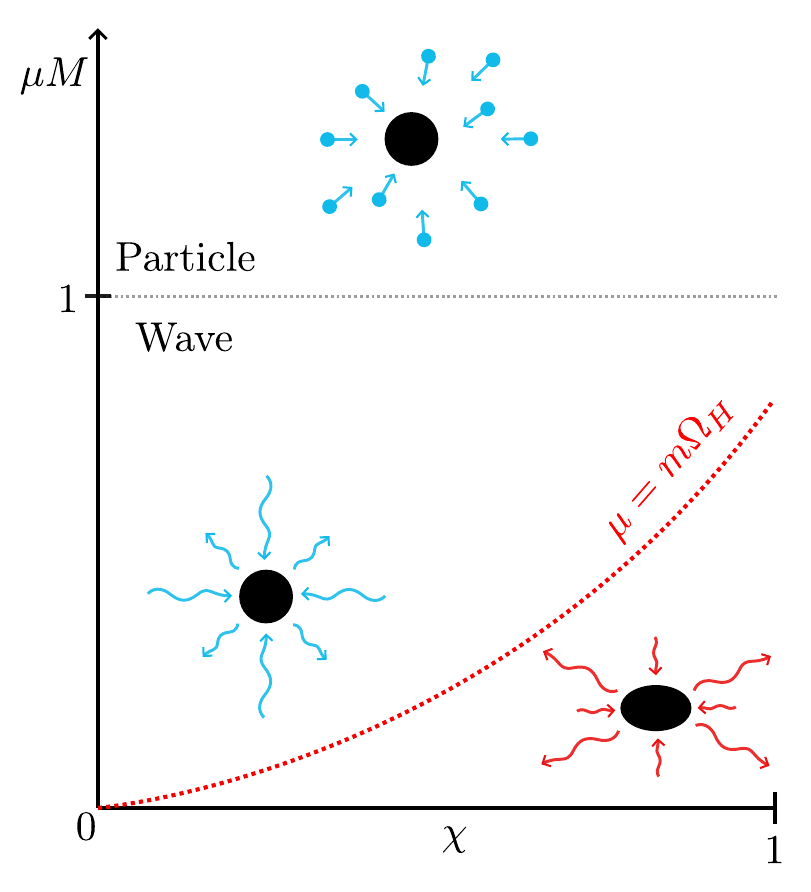}
    \caption{
    Diagram of the key regimes of vector dark matter surrounding a Kerr \bh{}, in the plane formed by the \bh{'s} dimensionless spin $\chi$ and mass parameter $\mu M$.
    For $\mu M \gtrsim 1$, the field appears to the \bh{} as a cloud of particles, and the accretion rate is maximized. For $\mu M < 1$, the field acts as a wave, and accretion is suppressed.
    If a mode satisfies the superradiant condition,
    then its ingoing waves are amplified by superradiant scattering, making the accretion rate negative, and therefore making the extraction rate positive.
    }
    \label{fig:accretion_regimes_cartoon}
\end{figure}

In this paper, we advance this project with an investigation, using the tools of \bh{} perturbation theory, into the environments of rotating \bh{s} surrounded by vector-field wave dark matter. This extends our previous study of the vector dark matter spike produced around a Schwarzschild \bh{}, published in Ref. \cite{Hancock:2025ois}, henceforth referred to as Paper I. 
There, we found that the problem split into two key regimes, identified with the dimensionless mass parameter $\mu M$, where $\mu$ and $M$ are the field mass parameter and the \bh{} mass, respectively.
For $\mu M \gtrsim 1$, the field acts as a cloud of particles, maximizing the mass accretion rate, while for $\mu M < 1$, the field acts as a wave, suppressing the accretion rate.
In this work, we find that an additional regime emerges within the wave regime when the \bh{} angular velocity $\Omega_H$ is nonzero. 
Decomposing the field in spherical harmonics $Y_\ell^m$, if a mode in this expansion satisfies the condition $\mu < m \Omega_H$, then its 
``accretion rate" $\dot{M}$ becomes negative, meaning its ``extraction rate" $-\dot{M}$ becomes positive, indicating the presence of superradiant scattering. 
The different regimes are shown schematically as regions in the $(\chi,\mu M)$ plane, 
where 
$\chi$
is the \bh{'s} dimensionless spin parameter, in Figure \ref{fig:accretion_regimes_cartoon}.

The study of vector \bh{} perturbations is quite advanced. As with \bh{} perturbations of any spin, it has hinged on the discoveries of various separation schemes---methods to separate the perturbation equations into a set of \ode{s}, thereby reducing the complexity of the problem drastically. 
For massless bosonic perturbations, Newman \& Penrose opened the door to general separability with their null-tetrad formalism~\cite{Newman:1961qr}. 
Using this, Price established separability for electromagnetic perturbations of the Schwarzschild spacetime~\cite{Price:1972pw}, and Teukolsky immediately followed with the same for Kerr~\cite{Teukolsky:1972my, Teukolsky:1973ha}. 
With both schemes fully capturing the (massless) field's dynamics, these effectively closed the methodological book for massless vector perturbations of stationary vacuum \bh{s}.
However, the question of separability remained open far longer for the Proca field, a massive vector field,
where the Newman-Penrose formalism does not lead to separated equations. 
Rosa \& Dolan~\cite{Rosa:2011my} found partial separation of the Proca equation in the Schwarzschild spacetime using a \vsh{} scheme, which was adapted to Kerr perturbations in Refs.~\cite{Pani:2012bp, Pani:2012vp, Pani:2013pma} 
for small \bh{} spin. 
Drawing on work by Lunin~\cite{Lunin:2017drx}, full separability was established by Frolov, Krtou\v{s}, Kubiz\v{n}\'{a}k, and Santos~\cite{Krtous:2018bvk, Frolov:2018ezx} who, using techniques from the study of \bh{s'} hidden symmetries~\cite{Frolov:2017kze}, found a separation scheme based on the principal tensor valid for all Kerr-NUT-(A)dS spacetimes.
Initially thought to only recover two of the field's three polarizations, it was later shown to capture all three of them
in the Kerr spacetime~\cite{Dolan:2018dqv}. 
Thus, full separability has been established for both massive and massless linear vector perturbations of both Schwarzschild and Kerr \bh{s}. 


Aside from immediate questions like the shifting of \qnm{s} from their Schwarzschild values, a key reason to explore perturbations of rotating \bh{s} has been to study rotational superradiance, or the amplification of incident waves via angular momentum extraction~\cite{Bekenstein:1998nt, Brito:2015oca}. 
This phenomenon was first investigated by Zel'dovich, who found that a body rotating at angular velocity $\Omega$ can amplify an incident wave of frequency $\omega$ if the condition
\begin{equation}
    \label{eq:superrad_condition}
    \omega < m \Omega\, 
\end{equation}
is satisfied~\cite{Zeldovich:1971ffh}.
In this paper, he pointed out that the phenomenon should apply for Kerr \bh{s}, also indicated by Misner~\cite{Misner:1972kx}. 
This Kerr \bh{} scenario 
was realized soon after in the classic discussions of Press \& Teukolksy~\cite{Press:1972zz} and Zel'dovich~\cite{Zeldovich:1972zqp}, for scalar and electromagnetic waves, respectively, and extended to gravitational waves in Ref.~\cite{Teukolsky:1974yv}. 

With these calculations came also the proposal of the ``\bh{} bomb,''
or superradiant instability
~\cite{Press:1972zz, Zeldovich:1972zqp}, where a sufficiently reflective surface placed around the \bh{} could cause repeated amplifications upon repeated scatterings of a wave, thus creating an unstable runaway energy buildup.
In the case of a massive field, this ``mirror" appears in a natural way, with the mass creating a reflective barrier in the field's effective potential
for modes with $\omega\lesssim\mu$.
Following this proposal, unstable growing modes were soon found for a massive scalar field in the Kerr spacetime~\cite{Damour:1976kh, Detweiler:1980uk, Dolan:2007mj, Shlapentokh-Rothman:2013ysa}.
For the massive vector field, the instability was demonstrated explicitly in Refs.~\cite{Pani:2012vp, Pani:2012bp} in a small-spin expansion. 
Calculations of this type were sufficient to study the energy-level structure~\cite{Baumann:2019ztm} and standard-model interactions~\cite{Siemonsen:2022ivj} of the resulting superradiant clouds,
and to place initial \bh{} superradiance constraints on ultralight vector dark matter using known astrophysical \bh{s}~\cite{Baryakhtar:2017ngi, Tsukada:2020lgt}. 
These studies also motivated the construction of stationary \bh{} solutions with a nontrivial bosonic field buildup in the full nonperturbative regime---i.e., hairy \bh{s}---done by placing the system directly at the superradiance threshold~\cite{Herdeiro:2014goa, Herdeiro:2016tmi}.

Parallel to these developments,
massive advances were made in the time domain. 
Since the Proca-field superradiant instability has a much shorter timescale than its scalar counterpart, robust numerical simulations of the former are far less expensive than the latter;
indeed, full 3+1-dimensional simulations of the instability are currently tractable in the Proca-field case but not in the scalar-field case. 
These evolutions of the Proca instability were done, first for a Proca test field~\cite{Witek:2012tr, East:2017mrj}, then with full backreaction, showing the system's evolution from the instability's activation until its ``quenching" after sufficiently slowing the \bh{'s} rotation~\cite{East:2017ovw, East:2018glu}. Finally, with the Proca separation scheme of Refs.~\cite{Krtous:2018bvk, Frolov:2018ezx}, the instability growth rates were calculated with no restrictions on \bh{} spin in Ref.~\cite{Dolan:2018dqv}. 

The present work joins this long line of study---Proca-field perturbations of rotating spacetimes---with a recent area of inquiry:
\bh{s} accreting from an infinite bath of wave dark matter.
Wave dark matter refers to a bosonic field of mass $\mu \lesssim 10\, \ev$ with a de Broglie wavelength exceeding its average inter-particle spacing and, thus, making it well described by a classical field~\cite{Hui:2021tkt}.
If populated as a condensate, with time dependence $e^{-i \mu t}$, such a field behaves as cold dark matter~\cite{Antypas:2022asj}. For massive vector, or ``dark photon", dark matter, many viable mechanisms to generate such a condensate exist, such as misalignment~\cite{Arias:2012az, Nelson:2011sf} and direct production during inflation~\cite{Graham:2015rva}. 

Recent work has studied how \bh{s} are affected in a universe filled with a 
scalar wave dark matter condensate. 
To model this condensate,
one is forced to relax the usual field falloff conditions far from the \bh{,} assuming it to instead reach a constant density, 
and thus bypassing the no-boson-hair theorems of Bekenstein~\cite{Bekenstein:1971hc, Bekenstein:1972ky, Bekenstein:1972ny}. 

This program began by modeling a single nonrotating \bh{} accreting from a scalar condensate, calculating the density profile, accretion rate, and other properties of the resulting dark matter ``spike", both numerically~\cite{Clough:2019jpm} and in linear perturbation theory~\cite{Hui:2019aqm}.
Both were soon extended to a rotating \bh{}~\cite{Bamber:2020bpu}, opening the door to a study of how wave dark matter accretion mixes with
superradiant growth in a quasi-adiabatic approximation~\cite{Hui:2022sri}. 
The numerical program has now been extended to simulations of binary \bh{}~\cite{Bamber:2022pbs, Aurrekoetxea:2023jwk, Aurrekoetxea:2024cqd, Cheng:2025wac} and binary neutron star~\cite{Srikanth:2025lic} mergers, finding characteristic dephasing of the mergers' \GW{} signals, both from 
dynamical friction and from wave-specific effects.
Results of these simulations were recently used to find tentative
evidence of a wave dark matter environment in LIGO-Virgo-KAGRA \GW{} data for the binary \bh{} merger GW190723~\cite{Roy:2025qaa}.

As is often the case, the scalar-field program 
has progressed well ahead of its vector counterpart. 
We initiated the latter
with a perturbative study of a Schwarzschild \bh{} accreting from a vector dark matter condensate, published in Paper I of this series~\cite{Hancock:2025ois}.
Using a composite of the \vsh{} and \lfkks{} approaches, an approach which has also been used to study reflection and transmission coefficients~\cite{Vispute:2026vek}, we recovered all Proca polarizations and showed that they all form a spike if connected to the bath.
Reducing largely to the same behavior as the scalar case in the ``particle" regime, where the field's Compton wavelength satisfies $\lambda_c \sim 1/\mu \lesssim M$, with \bh{} mass $M$ in natural units, the results diverge in the ``wave" regime, with $\lambda_c > M$.
In particular, the vector's monopole mode and fundamental 
$s=-1$ mode had wave-regime accretion rates two to three orders of magnitude larger than both the next largest vector-field mode and the scalar.
This mirrors the 
behavior of the vector field's Schwarzschild quasi-bound states, where the $s=-1$ polarization decays faster by a similar factor at the same angular momentum $\ell$ than both the other vector polarizations and its scalar counterpart~\cite{Rosa:2011my}. 
These results thus give a key indication of the main differences between the vector and scalar cases, a thread to be pulled on with follow-up work.

In this paper, we extend our previous work in Paper I to a rotating spacetime, showing that the basic shape of the dark matter spike persists and exploring how the accretion rates are changed by the presence of a superradiant regime.
As in Paper I, we fix the field's frequency $\omega$ to its mass, $\omega = \mu$,
modeling an equilibrium with an asymptotic dark matter bath.
That is, we fix the frequency to a real number, a different approach from 
most \bh{} perturbation theory studies, which determine $\omega$ as a complex eigenvalue. 

We employ the \lfkks{} scheme~\cite{Krtous:2018bvk, Frolov:2018ezx} to separate the equations, constructing solutions for the monopole and the fundamental $m=1$ mode for each of the three polarizations. 
We find that both the \bh{'s} mass and angular momentum accretion rates tend to zero at the superradiance threshold, given in Eq.~\eqref{eq:superrad_condition}, beyond which they become negative, thus indicating energy extraction from the \bh{} by the field. 
We finish by comparing these results with those of the scalar-field case and deriving the relevant timescales for comparisons, in particular comparing that of \bh{} mass accretion with the superradiant instability timescale.

This paper is structured as follows. In Section \ref{sec:setup}, we introduce the Proca field as a perturbation of the Kerr spacetime, and discuss its basic properties. In Section \ref{sec:eq_separating}, we separate it into an angular and a radial equation using the \lfkks{} scheme. We then solve these equations in Sections \ref{sec:angular_eq_solving} and \ref{sec:radial_eq_solving}, using a harmonic procedure for the former and a numerical procedure for the latter. In Section \ref{sec:results}, we use our solutions to compute the field's density profile around the \bh{} and the rates at which it deposits mass and angular momentum into the horizon. We conclude with a discussion of these results, their relevance to the literature, and future work in Section \ref{sec:discussion}.

\paragraph*{A note on notation:}
In this work, we
adopt a $(-,+,+,+)$ metric signature and use natural units $c=G=\hbar=1$ throughout, except when explicitly stated otherwise.
For (anti-) symmetrization
of indices, we use the standard parenthesis/bracket shorthand notation, i.e.,
\begin{align}
    T_{[\mu\nu]} = {}&\frac{1}{2!}\left(T_{\mu\nu} - T_{\nu\mu}\right)
\,,\quad
    T_{(\mu\nu)} = \frac{1}{2!}\left(T_{\mu\nu} + T_{\nu\mu}\right)\, .
\end{align}
Tensors
may be written in ``index-free" notation.
One may expand these in a coordinate basis of appropriate rank to restore indices,
e.g., $v = v^\mu\partial_\mu$. We will use Greek indices $\{\mu, \nu, \ldots\} \in \{0,\ldots, 3\}$ for components of spacetime objects and Latin indices $\{i, j, \ldots\} \in \{1,\ldots, 3\}$ for components of objects on spatial slices. 
For single-argument functions, we may use primes to denote derivatives with respect to the argument, e.g., $f'(x)\coloneq \frac{\dif}{\dif x} f(x)$.
We variously characterize the \bh{'s} spin by $\chi$, the dimensionless spin; $a$, the spin parameter; $J$, the angular momentum; and $\Omega_H$, the horizon angular velocity. These are related by $\chi = a/M = J/M^2$ and $\Omega_H = \frac{a}{2 M r_+}$, where $r_+ = M+\sqrt{M^2-a^2}$ is the outer horizon radius.
When speaking of a complex field frequency $\omega$, we often split the field into its real and imaginary parts, notated by $\omega = \omega_R + i \omega_I$.

\section{Setup}
\label{sec:setup}

\subsection{Action and field equations}

To begin, we consider a $4$-dimensional Lorentzian manifold $\mathcal{M}$, equipped with a Lorentzian metric $g_{\mu\nu}$ which is  minimally coupled to a complex\footnote{Realistic vector dark matter is generally modeled as a real field. However, the complexity of $A_\mu$ has no effect on the results in this work, so we make it complex for computational convenience.} Proca field $A_\mu$ of with mass parameter $\mu$.
We associate the latter with a field-strength tensor
\begin{align}
\label{eq:FmunuForm}
F_{\mu\nu} & \coloneq (\dif A)_{\mu\nu}
    = \nabla_{\mu} A_{\nu} - \nabla_{\nu} A_{\mu}
\,.
\end{align}
Assuming
Einstein-Hilbert dynamics for
the metric $g_{\mu\nu}$,
the Einstein-Proca action takes the form
\begin{align}
\label{eq:ActionEinsteinProca}
S & = \int_{\mathcal{M}} \dif^{4}x \sqrt{-g} \left(
    \frac{R}{2 \kappa} + \Lie_{\rm{Proca}}
\right)
\,,
\end{align}
where
$R$ is the Ricci scalar associated with $g_{\mu\nu}$
and the gravitational coupling is $\kappa=8\pi = M_{\text{pl}}^{-2}$, where $M_{\text{pl}}$ is the Planck mass.
The Proca Lagrangian is given by
\begin{align}
\label{eq:ProcaLagrangian}
\Lie_{\rm{Proca}} & =
    -\frac{1}{4} F^{\ast}_{\mu\nu}F^{\mu\nu} - \frac{\mu^2}{2}A^{\ast}_\mu A^\mu
\,.
\end{align}
From the matter sector of the Einstein--Proca action in Eq.~\eqref{eq:ActionEinsteinProca},
we find the energy-momentum tensor
\begin{align}
\label{eq:ProcaTmunu}
T_{\mu\nu} & =
    F^\ast_{(\mu}{}^{\rho} F^{}_{\nu)\rho}
+ \mu^2 A^{\ast}_{(\mu} A^{}_{\nu)} + g_{\mu\nu} \Lie_{\rm{Proca}}
\,.
\end{align}
Extremizing the action in Eq.~\eqref{eq:ActionEinsteinProca}
with respect to the vector field and the metric
yields the Proca equations and the Einstein equations, respectively, 
\begin{subequations}
\begin{align}
\label{eq:ProcaEoM}
\nabla_{\nu} F^{\mu\nu} + \mu^2 A^{\mu} & = 0\\
\label{eq:einstein_eqs}
R_{\mu\nu}-\frac{1}{2}R g_{\mu\nu} &= \kappa T_{\mu\nu}
\,.
\end{align}
\end{subequations}
In the limit of vanishing mass, $\mu \to0$, Eq. \eqref{eq:ProcaEoM} reduces to Maxwell's equation
and we are free
to choose a gauge for the vector field.
This gauge freedom is lost for a nonvanishing mass, $\mu\neq0$,
and the Lorenz condition,
\begin{align}
    \label{eq:Lorenzgauge}
    \nabla_{\mu}A^{\mu} = {}&0
    \, ,
\end{align}
must be satisfied.
To see this, one may take a covariant derivative of the Proca equation~\eqref{eq:ProcaEoM} and use the identity $\nabla_{\mu} \nabla_{\nu} F^{\mu\nu} = 0$.

\subsection{Kerr background}

We are interested in a single astrophysical \bh{} far from any other compact bodies. The Kerr spacetime is a worthy approximation, describing a stationary rotating \bh{} in vacuum. In Boyer-Lindquist coordinates $(t,r,\theta,\phi)$, its metric is
\begin{align}
\label{eq:kerr_metric}
\dif s^2 & = -\left(1 - \frac{2 M r}{\Sigma}\right) \dif t^2 
    - \frac{4 M a r \sin^2\theta}{\Sigma} \dif t \dif \phi 
    + \frac{\Sigma}{\Delta} \dif r^2
\nonumber\\  &
    + \Sigma \dif \theta^2 
    + \left[r^2 + a^2 + \frac{2 M a^2 r \sin^2 \theta}{\Sigma}\right] \sin^2 \theta \dif \phi^2
\,,   
\end{align}
where
\begin{subequations}
\label{eq:KerrMetricFunctions}
\begin{align}
    \Sigma \coloneq{}& r^2+a^2 \cos^2\theta\\
    \label{eq:Delta_def}
    \Delta \coloneq{}& r^2 - 2 M r + a^2 = (r-r_+)(r-r_-)\\
    r_\pm \coloneq{}& M \pm \sqrt{M^2-a^2}\, .
\end{align}
\end{subequations}
The parameters $M \in \mathbb{R}^+$ and $a \in [0,M)$ may be interpreted as the \bh{'s} mass and spin, respectively. In the non-rotating limit, $a \to 0$, Eq. \eqref{eq:kerr_metric} reduces to the Schwarzschild metric.

\subsubsection{Comment on test-field approximation}

We will treat the Proca field $A_\mu$ as a test field, propagating on the Kerr spacetime without influencing it. In Appendix~D of Ref.~\cite{Hancock:2025ois}, we showed this approximation to be valid for a Schwarzschild \bh{} of virtually any astrophysical mass, $5\, \msun \lesssim M \lesssim 10^{10}\, \msun$, given the ambient dark matter density $\rho_c$ is no higher than about $\rho_c \sim 1\, \msun/\pc^3 \sim 10\, \Gev/\cm^3$, 
a density appropriate to the core of a dark matter soliton in a large galaxy~\cite{Schive:2014hza}.
For this analysis, we compared the Ricci curvature induced by the field to the square root of the background spacetime's Kretschmann scalar, finding the latter to grow much faster than the former near the horizon, and therefore that backreaction matters least near the \bh{} horizon.
Dark matter backreaction effects are most important near the edge of the \bh{} sphere of influence;
for the largest \bh{s}, of mass $M \sim 10^{10}\, \msun$, the two curvature scalars are comparable in magnitude at $r\sim 10^6 M$ for $\rho_c \sim 10\, \msun/\pc^3$, meaning backreaction can not be neglected in this region.
However, the field's density must be increased by several more orders of magnitude before backreaction becomes relevant near the horizon for supermassive \bh{s}, or at all for stellar-mass \bh{s}. 


Those validity arguments are not significantly modified by the inclusion of \bh{} spin, so these bounds persist in this work. 
When required to make a choice, we will generally quote figures for a
\bh{} of mass $M=10^9\, \msun$ in a 
dark matter environment with density
$\rho_c = 10\, \msun/\pc^3$. 
While near the upper end of the bounds, this approximation remains valid for the \bh{'s} entire sphere of influence in such a system, to our degree of precision. 

\subsection{Polarizations and nomenclature}

The Proca equation \eqref{eq:ProcaEoM} is a system of four coupled linear \pde{s}. By standard degree-of-freedom counting arguments, one can show that it propagates three physical polarizations. As is standard in the literature, we will label each of these by its ``spin projection" 
$s\in \{-1,0,1\}$~\cite{Galtsov:1984ixy, Rosa:2011my}. Each projection enjoys a definite behavior under parity inversion $(\theta,\phi) \mapsto (\pi-\theta,\phi+\pi)$, with the $s=0$ mode being ``odd" and the $s=\pm 1$ modes being ``even".
In the Schwarzschild limit $a\to 0$, the \vsh{} approach decouples the Proca equation into two sectors: an odd-parity sector containing the $s=0$ polarization and an even-parity sector containing the $s=\pm1$ polarizations.
As we will see, in the \lfkks{} approach needed to separate the equations in the Kerr spacetime, this manifest association is broken, and we will generally identify the spin projection
$s$ of solutions using their behavior in the $a \to 0$ limit. 
For a detailed summary of the nomenclature and properties of these polarizations, we refer readers to Table 1 of Paper~1~\cite{Hancock:2025ois}. 

\section{Separated equations}
\label{sec:eq_separating}

Unlike its Minkowski counterpart, the Proca equation~\eqref{eq:ProcaEoM} in the Kerr spacetime does not in general admit exact solutions, nor does its form suggest obvious approximation schemes. Given its linear structure, however, we should expect it to inherit the symmetry structure of the background spacetime; that is, it should admit towers of solutions which are eigenfunctions of the Lie derivatives $\Lie_t$ and $\Lie_\phi$ along the isometry generators $\partial_t$ and $\partial_\phi$. Moreover, we may draw hope from the analogous massive scalar model, which fully separates into a system of \ode{s} in the Kerr spacetime, one for $r$ and one for $\theta$~\cite{Damour:1976kh, Detweiler:1980uk, Dolan:2007mj}. Here, however, it is unclear what ansatz one would take for the field $A_\mu$ to search for such behavior. This was an open question until the pioneering work of Lunin, Frolov, Krtou\v{s}, Kubiz\v{n}\'{a}k, and Santos (LFKKS), who, following a detailed study of spacetime hidden symmetries, found a highly general separation scheme for Proca and electromagnetic fields, valid in all spacetimes satisfying certain symmetry properties, including Kerr~\cite{Krtous:2018bvk, Frolov:2018ezx}. We use their scheme in this work, and we summarize it here for completeness.

\subsection{Lunin-Frolov-Krtou\v{s}-Kubiz\v{n}\'{a}k-Santos scheme}
\label{sec:fkks}

The \lfkks approach~\cite{Krtous:2018bvk, Frolov:2018ezx} reduces the Proca equation to a set of decoupled \ode{s} for any background spacetime in the Kerr-NUT-(A)dS class.
It has been used extensively to study both quasi-normal modes and quasi-bound states for Proca perturbations of the Kerr spacetime~\cite{Dolan:2018dqv,Dolan:2019hcw,Frolov:2018ezx, Percival:2020skc, Siemonsen:2019ebd}. 
The approach rests on an object known as the principal tensor, a non-degenerate closed conformal Killing-Yano 2-form, which exists only for certain spacetimes. 
For example,
the Schwarzschild and Schwarzschild-(A)dS spacetimes
admit only a degenerate closed conformal Killing-Yano $2$-form rather than a true principal tensor. 
The \lfkks{} method has been shown to separate the Proca equation in them, though it recovers only the even-parity ($s=\pm1$) modes~\cite{Fernandes:2021qvr, Hancock:2025ois}.
The Kerr spacetime, conversely, admits a true principal tensor, and is thus an ideal case for the \lfkks{} scheme.
While the approach was originally thought to only recover the even-parity ($s=\pm1$) polarizations in this case, the odd-parity ($s=0$) polarization was soon shown to also be present~\cite{Dolan:2018dqv}. 
We can thus treat all three polarizations with the \lfkks{} scheme in this work. 

In the Kerr spacetime,
the principal tensor in Boyer-Lindquist coordinates takes the form~\cite{Frolov:2017kze}
\begin{align}
\label{eq:princ_tens_bl_coords}
h = {} & r \dif t \wedge \dif r
    + a r \sin^2\theta\, \dif r\wedge \dif \phi
\\{}&
    - a\cos\theta \sin\theta\, \dif \theta \wedge \left(a \dif t - \left(a^2 + r^2\right) \dif \phi\right)
\,.\nonumber   
\end{align}
We may separate the Proca equation using the ansatz~\cite{Krtous:2018bvk}
\begin{equation}
    \label{eq:fkks_ansatz}
    A^\mu = B^{\mu\nu}\nabla_\nu Z\, ,
\end{equation}
where $Z$ is an undetermined scalar function and $B^{\mu\nu}$ is the polarization tensor.
The latter is determined by
\begin{equation}
    \label{eq:polariz_tensor_eq}
    B^{\mu\rho}\left(g_{\rho\nu} + i\nu h_{\rho\nu}\right) = \delta^\mu_\nu\, ,
\end{equation}
where $\nu$ is an undetermined separation constant.
Adopting Eq.~\eqref{eq:fkks_ansatz} reduces Eq.~\eqref{eq:ProcaEoM} to a single \pde{} for the scalar $Z$. To see this, we borrow a result from Ref.~\cite{Krtous:2018bvk} that
\begin{align}
    \nabla_\nu F^{\mu\nu} = -B^{\mu\nu}\nabla_\nu \left(\nabla_\rho \nabla^\rho Z - 2 i \nu\, t^\rho A_\rho\right)\, ,
    \label{eq:proca_fkks}
\end{align}
\pagebreak
where $t^\mu\partial_\mu = \partial_t$ is the timelike Killing vector. The Proca equation (\ref{eq:ProcaEoM}) is therefore satisfied if
\begin{equation}
    \label{eq:Z_eq}
    \left(\nabla_\mu \nabla^\mu - 2 i \nu\, t_\mu B^{\mu\nu} \nabla_\nu - \mu^2\right) Z = 0\, .
\end{equation}
Thus, the Proca equation is reduced to a wave-type equation for the scalar $Z$.

\subsection{Separation for Kerr spacetime}

We now restrict our calculation to the Kerr spacetime. Solving Eq.~\eqref{eq:polariz_tensor_eq} using the Kerr principal tensor in Boyer-Lindquist coordinates, Eq. \eqref{eq:princ_tens_bl_coords}, we find the polarization tensor expression
\begin{widetext}
\begin{align}
    \label{eq:polarization_tens_kerr}
    {}&B^{\mu\nu} = \left(
    \begin{array}{>{\displaystyle}c>{\displaystyle}c>{\displaystyle}c>{\displaystyle}c}
     \frac{1}{\Sigma}\left(\frac{a^2 \sin^2{\theta}}{q_\theta}-\frac{\left(a^2+r^2\right)^2}{q_r \Delta}\right) {}& \frac{i \nu  r \left(a^2+r^2\right)}{q_r \Sigma} {}& \frac{i a^2 \nu  \sin{\theta} \cos{\theta}}{q_\theta \Sigma} {}& \frac{a}{\Sigma} \left(\frac{1}{q_\theta}-\frac{a^2+r^2}{q_r \Delta}\right) \\[13pt]
     -\frac{i \nu  r \left(a^2+r^2\right)}{q_r \Sigma} {}& \frac{\Delta}{q_r \Sigma} {}& 0 {}& -\frac{i a \nu  r}{q_r \Sigma} \\[13pt]
     -\frac{i a^2 \nu  \sin{\theta} \cos{\theta}}{q_\theta \Sigma} {}& 0 {}& \frac{1}{q_\theta \Sigma} {}& -\frac{i a \nu  \cot{\theta}}{q_\theta \Sigma} \\[13pt]
     \frac{a}{\Sigma} \left(\frac{1}{q_\theta}-\frac{a^2+r^2}{q_r \Delta}\right) {}& \frac{i a \nu  r}{q_r \Sigma} {}& \frac{i a \nu  \cot{\theta}}{q_\theta \Sigma} {}& \frac{1}{\Sigma}\left(\frac{\csc^2{\theta}}{q_\theta}-\frac{a^2}{q_r \Delta}\right) \\
    \end{array}
    \right) \, ,
\end{align}
\end{widetext}
where we introduce
\begin{align}
    q_r \coloneq 1 + \nu^2 r^2\, , \quad q_\theta \coloneq 1 - a^2\nu^2 \cos^2\theta
\,,
\end{align}
and recall that $\nu$ is the (undetermined) separation constant in Eq.~\eqref{eq:polariz_tensor_eq}.
Adopting Eq.~\eqref{eq:polarization_tens_kerr}, the equation for the scalar Z, Eq.~\eqref{eq:Z_eq}, becomes a lengthy \pde{}. 
It is, however, separable under the single-mode ansatz
\begin{equation}
\label{eq:fkks_Z_ansatz}
    Z = R(r) S(\theta) e^{-i \omega t} e^{i m \phi}
\, ,
\end{equation}
where $\omega$ and $m$ are eigenvalues corresponding with the spacetime's isometries, generated by $\partial_t$ and $\partial_\phi$, respectively. Requiring that solutions be periodic in $\phi$ fixes $m \in \mathbb{Z}$, while the allowable values of $\omega$ are fixed by the physical situation under consideration. 

Imposing Eq.~\eqref{eq:fkks_Z_ansatz}, one finds that Eq.~\eqref{eq:Z_eq} separates into two \ode{s}: 
a radial equation for $R(r)$ and an angular equation for $S(\theta)$,
\begin{subequations}
\begin{align}
\label{eq:kerr_proca_radial_fkks}
0 = {}& q_r \frac{d}{dr} \left( \frac{\Delta}{q_r} R '(r)\right) 
    + \left(\frac{K_r^2}{\Delta} -\frac{2 \nu  K_r}{q_r} - \mu^2 r^2 - \kappa_1\right) R (r)   
\\ 
    0 = {}& \frac{q_\theta}{\sin \theta} \frac{d}{d\theta} \left(\frac{\sin \theta}{q_\theta} S'(\theta)\right)
\nonumber\\ {}&
    + \biggl((q_\theta-1) \left[\frac{m^2}{a^2 \nu^2 \sin^2\theta} + \frac{\mu^2 - \omega^2}{\nu}\right]
\nonumber \\ {}& \qquad
    - \frac{2}{q_\theta}\frac{\sigma}{\nu} 
    + \frac{2 \omega}{\nu} 
    - (m - a \omega)^2 + \kappa_1 \biggr) S(\theta) 
    \label{eq:kerr_proca_angular_fkks}\, ,
\end{align}
\end{subequations}
with separation constant $\kappa_1$, 
where we recall that primes denote derivatives with respect to the function's argument. 
The auxiliary variables are
\begin{alignat}{2}
    \nonumber
    K_r = {}&(a^2 + r^2)\omega - am\, , \quad {}&K_\theta ={}& m - a \omega \sin^2\theta\\
     \sigma ={}& \omega + a \nu^2(m - a \omega)\, .
    \label{eq:aux_vars_defs}
\end{alignat}
Under the same treatment, the Lorenz condition in Eq.~\eqref{eq:Lorenzgauge} yields the separated equations
\begin{subequations}
\begin{align}
    0 = {}&\frac{d}{dr} \left(\frac{\Delta}{q_r} R'(r)\right)+ \left(\frac{K_r^2}{\Delta q_r} + \frac{2-q_r}{q_r^2} \frac{\sigma}{\nu} + \kappa_2 + \frac{\omega}{\nu} \right)R(r)
    \label{eq:kerr_lorenz_radial_fkks}\\\nonumber
    0 = {}& \frac{1}{\sin \theta} \frac{d}{d\theta} \left(\frac{\sin \theta}{q_\theta} S'(\theta)\right)\\
    {}& - \left(\frac{K_\theta^2}{q_\theta \sin^2 \theta} + \frac{2 - q_\theta}{q_\theta^2} \frac{\sigma}{\nu} - \kappa_2 - \frac{\omega}{\nu} \right) S(\theta)
    \label{eq:kerr_lorenz_angular_fkks} \, ,
\end{align}
\end{subequations}
with additional separation constant $\kappa_2$.
Subtracting Eq.~\eqref{eq:kerr_lorenz_radial_fkks}, multiplied by $q_r$, from Eq.~\eqref{eq:kerr_proca_radial_fkks}
leads to the characteristic equation
\begin{equation}
    \kappa_2 - \kappa_1 + a \nu (m - a \omega) + r^2 \nu^2\left(\kappa_2 - \frac{\mu^2}{\nu^2} + \frac{\omega}{\nu}\right) = 0\, ,
\end{equation}
from which we may read off the separation constants
\begin{subequations}
\begin{align}
    \kappa_2 = {}&\frac{\mu^2}{\nu^2} - \frac{\omega}{\nu}\\
    \kappa_1 = {}& \frac{\mu^2}{\nu^2} - \frac{\omega}{\nu} + a \nu (m - a \omega)\, .
\end{align}
\end{subequations}
With these fixed, the Proca equation and Lorenz condition now identically give the separated equations
\begin{subequations}
\label{eq:fkks_eq_final}
\begin{align}
    \label{eq:fkks_radial_eq_final}
    0 = {}&\frac{d}{dr}\left(\frac{\Delta}{q_r} R'(r)\right) + \left(\frac{K_r^2}{q_r \Delta} + \frac{2 - q_r}{q_r^2} \frac{\sigma}{\nu} - \frac{\mu^2}{\nu^2}\right)R(r)\\\nonumber
    \label{eq:fkks_angular_eq_final}
    0 = {}&\frac{1}{\sin \theta} \frac{d}{d\theta} \left(\frac{\sin \theta}{q_\theta} S'(\theta) \right)\\
    {}&- \left(\frac{K_\theta^2}{q_\theta \sin^2\theta} + \frac{2 - q_\theta}{q_\theta^2} \frac{\sigma}{\nu} - \frac{\mu^2}{\nu^2} \right) S(\theta) \, .
\end{align}
\end{subequations}
From the solutions to Eqs.~\eqref{eq:fkks_eq_final}, we may reconstruct the scalar $Z$ using Eq.~\eqref{eq:fkks_Z_ansatz}, 
then reconstruct the vector-field solutions, $A^{\mu}$, to the Proca equation~\eqref{eq:ProcaEoM}, using Eq.~\eqref{eq:fkks_ansatz}.

\section{Solving the angular equation}
\label{sec:angular_eq_solving}

Unlike the Schwarzschild case, where the field's angular equation is solved by spherical harmonics, the Kerr angular equation~\eqref{eq:fkks_angular_eq_final} has no obvious basis of known solutions. 
Although it is similar in form 
to the well-known spin-weighted spheroidal harmonic equation---the angular Teukolsky equation~\cite{Teukolsky:1972my, Teukolsky:1973ha, Berti:2005gp}---it
inherits a more complicated structure from the  \lfkks{} equation~\eqref{eq:Z_eq} 
with features like a nonlinear dependence on the separation constant $\nu$ and a mass term. 
While we may solve it along similar lines to the angular Teukolsky equation, we must treat this additional structure with care. 


For small spin or low-frequency waves, one can solve the Teukolsky angular equation via an expansion in $a \omega$~\cite{Press:1973zz, Berti:2005gp}.
For massive fields, one can do the same for small $a \omega$ and $\mu M$, or
a related parameter~\cite{Detweiler:1980uk}. This was done for the Proca field FKKS angular equation in Ref.~\cite{Baumann:2019eav}. 
These approaches, while useful for obtaining analytic formulae, have the drawback of only applying in regimes where the expansion parameter is small.\footnote{Though Ref.~\cite{Press:1973zz} uses a ``continuation" method to numerically extend the angular eigenvalue out of the $a \omega \ll 1$ regime.} 
One can, of course, numerically integrate the angular equation and determine the eigenvalue via a shooting method, but stiff behavior near the poles makes the problem ill-suited to finite differencing~\cite{Press:1973zz}.
Beyond these limited methods, the most popular approach has been to expand the angular function in a complete basis of functions satisfying some useful property. Leaver's classic paper~\cite{Leaver:1985ax},
for instance, solves the angular Teukolsky equation using a basis of polynomial and exponential factors in $u=\cos\theta$ which satisfy the regularity boundary condition at $u=-1$. 
Ref.~\cite{Baumann:2019eav} does the same for the \lfkks{} angular equation, except with Chebyshev cardinal polynomials in $\cos\theta$.
Drawing inspiration from the Schwarzschild limit, one can also use (spin-weighted) spherical harmonics as a basis, finding the spheroidal harmonics as linear combinations of their spherical counterparts.
This is done for the angular Teukolsky equation using spin-weighted spherical harmonics in Refs.~\cite{Hughes:1999bq, Cook:2014cta} and for the \lfkks{} angular equation using standard spherical harmonics in Ref.~\cite{Dolan:2018dqv}. Since we are interested principally in reconstructing the angular equation for the field's most fundamental modes, we borrow the latter's approach.


\subsection{General procedure}

We begin by expanding the angular function $S(\theta)$ as
\begin{equation}
    \label{eq:S_expansion}
    S(\theta) = \sum_{\ell, m} b_\ell Y^m_\ell(\theta)\, ,
\end{equation}
where $b_\ell$ are constants, and $Y_\ell^m(\theta)$ is related to the spin-$0$ spherical harmonic by $Y^m_\ell(\theta, \phi) = Y^m_\ell(\theta) e^{i m \phi}$. Plugging Eq.~\eqref{eq:S_expansion} into the angular equation \eqref{eq:fkks_angular_eq_final} yields a lengthy expression involving first and second derivatives of $Y^m_\ell (\theta)$.
From the defining equation for the spherical harmonics $Y_\ell^m(\theta,\phi)$, one can derive the identity
\begin{equation}
    \label{eq:sph_harm_second_der}
    \frac{\dif^2}{\dif \theta^2}Y^m_\ell(\theta) = \cot\theta \frac{\dif}{\dif \theta}Y^m_\ell(\theta) - \left(\ell(\ell+1) - \frac{m^2}{\sin^2\theta}\right)Y^m_\ell(\theta)\, .
\end{equation}
This identity applies to $Y_\ell^m(\theta)$ at any $a$, and we may use it to eliminate the second-derivative terms Eq.~\eqref{eq:fkks_angular_eq_final}.
This procedure yields an equation of the form
\begin{align}
    \label{eq:Y_operator_eq}
    \hat{L}_\theta^{\ell}\, Y_\ell^m(\theta) = 0,
\end{align}
where $\hat{L}_\theta^\ell$ is a first-order differential operator given by
\begin{align}
    \hat{L}_\theta^\ell ={}& - 2 a^2 \nu ^2 \cos\theta\sin\theta\, \frac{\dif}{\dif \theta}\\\nonumber
    {}& + a^2 \left(k^2-\Lambda  \nu ^2-2 \nu  \sigma +\nu ^2 \ell  (\ell +1)\right) \cos^2 \theta\\\nonumber
    & - a^4 k^2 \nu ^2 \cos^4 \theta
    + \left(\Lambda - \ell(\ell+1)\right)\, .
\end{align}
Here, we introduce
\begin{subequations}
\begin{align}
    k^2 \coloneq {}& \sqrt{\omega^2-\mu^2}\\
    \Lambda \coloneq{}& \frac{\mu^2}{\nu ^2} - \frac{\sigma }{\nu } - a^2 \omega ^2+2 a m \omega\, .
\end{align}
\end{subequations}
Integrating Eq.~\eqref{eq:Y_operator_eq} against $\int d\theta \sin \theta \left(Y^m_{\ell'}(\theta)\right)^*$, 
and using the orthogonality relation for $Y_\ell^m(\theta)$,
we arrive at
\begin{align}
    \label{eq:angular_matrix_eq_full}
    0 ={}& \int \dif\theta \sin \theta \left(Y^m_{\ell'} (\theta)\right)^*\hat{L}_\theta^\ell\, Y_\ell^m(\theta)\\\nonumber
    \eqcolon{}& \sum_{\ell=|m|}^\infty b_\ell M^{\ell \ell'}
\end{align}
where the matrix $M^{\ell \ell'}$ is given by
\begin{align}
    \label{eq:angular_matrix_def}
    M^{\ell \ell'} = {}& a^2 \left(k^2-\Lambda  \nu ^2-2 \nu  \sigma +\nu ^2 \ell  (\ell +1)\right) \mathcal{C}_2^{\ell  \ell '}\\\nonumber
    &- a^4 k^2 \nu ^2 \mathcal{C}_4^{\ell  \ell '} - 2 a^2 \nu ^2 \mathcal{D}_2^{\ell  \ell '}+\left(\Lambda - \ell(\ell+1)\right) \delta ^{\ell  \ell '}\, .
\end{align}
Here, $\delta^{\ell \ell'}$ is the Kronecker delta, and the other coefficients are defined by
\begin{subequations}
\label{eq:trig_coeffs}
\begin{align}
    \label{eq:c2_coeff}
    \mathcal{C}_2^{\ell \ell'} \coloneq{}& \int_{S^2} \dif \Omega  \left(Y^m_{\ell'}\right)^* \cos^2 \theta\, Y_\ell^m\\
    \label{eq:c4_coeff}
    \mathcal{C}_4^{\ell \ell'} \coloneq{}& \int_{S^2} \dif \Omega  \left(Y^m_{\ell'}\right)^* \cos^4 \theta\, Y_\ell^m\\
    \label{eq:d2_coeff}
    \mathcal{D}_2^{\ell \ell'} \coloneq{}& \int_{S^2} \dif \Omega  \left(Y^m_{\ell'}\right)^* \cos \theta \sin\theta\, \partial_\theta Y_\ell^m\, .
\end{align}
\end{subequations}
We may find explicit closed-form expressions for the coefficients in Eqs. \eqref{eq:trig_coeffs} using the Gaunt coefficents, $G_{\ell_1, \ell_2, \ell_3}^{m_1, m_2, m_3}$,
which are closely related to the Clebsch-Gordan coefficients,
and defined by the integral
\begin{align}
    \label{eq:gaunt_coeff_def}
    G_{\ell_1, \ell_2, \ell_3}^{m_1, m_2, m_3}& \coloneq \int_{S^2} \dif \Omega  \left(Y_{\ell_1}^{m_1}\right)^* Y_{\ell_2}^{m_2} Y_{\ell_3}^{m_3}\\\nonumber
    = {}&\sqrt{\frac{\left(2 \ell _1+1\right) \left(2 \ell _2+1\right) \left(2 \ell _3+1\right)}{4\pi}} \\\nonumber
    &\times\left(
        \begin{array}{ccc}
             \ell_1 & \ell_2 & \ell_3 \\
             0 & 0 & 0 \\
        \end{array}
        \right) \left(
        \begin{array}{ccc}
             \ell_1 & \ell_2 & \ell_3 \\
             m_1 & m_2 & m_3 \\
        \end{array}
    \right)\, ,
\end{align}
where $\left(\begin{array}{ccc}
             \cdot & \cdot & \cdot \\
             \cdot & \cdot & \cdot \\
        \end{array}\right)$
is the Wigner 3-j symbol. 
To express the three coefficients in Eqs. \eqref{eq:trig_coeffs} in terms of Gaunt coefficients, one can recall the explicit expressions for the $\ell=0,2,4$ spherical harmonics,
and then reconstruct the trigonometric factors in the integrands in terms of these spherical harmonics. 
With the resulting integrals, and recalling the standard expression for $\partial_\theta Y_\ell^m$, 
one can use Eq.~\eqref{eq:gaunt_coeff_def} to find
\begin{subequations}
\begin{align}
    \mathcal{C}_2^{\ell \ell'} ={}& \frac{2\sqrt{\pi }}{3}  G_{{\ell ,0,\ell '}}^{{m,0,m}}+\frac{4}{3} \sqrt{\frac{\pi }{5}} G_{{\ell ,2,\ell '}}^{{m,0,m}}\\
    \mathcal{C}_4^{\ell \ell'} ={}& \frac{2\sqrt{\pi }}{5}  G_{{\ell ,0,\ell '}}^{{m,0,m}}+\frac{8}{7} \sqrt{\frac{\pi }{5}} G_{{\ell ,2,\ell '}}^{{m,0,m}}+\frac{16 \sqrt{\pi }}{105}  G_{{\ell ,4,\ell '}}^{{m,0,m}}\\
    \mathcal{D}_2^{\ell \ell'} ={}& \frac{2\sqrt{\pi }}{3}  m\, G_{{\ell ,0,\ell '}}^{{m,0,m}}+\frac{4}{3} \sqrt{\frac{\pi }{5}} m\, G_{{\ell ,2,\ell '}}^{{m,0,m}}\\\nonumber
    {}&+ 2 \sqrt{\frac{2 \pi }{15}} \sqrt{({\ell '}-m) (m+{\ell '}+1)}\, G_{{\ell ,2,\ell '}}^{{m,-1,1 + m}}\, .
\end{align}
\end{subequations}
We have now reduced the angular equation~\eqref{eq:fkks_angular_eq_final} to a matrix equation~\eqref{eq:angular_matrix_eq_full} for the constants $b_\ell$. 
Inspecting the structure of the matrix $M^{\ell \ell'}$ in Eq.~\eqref{eq:angular_matrix_def}, one notices that
that for a given $m$,
odd-$\ell$ modes are coupled only to odd-$\ell$ modes, and likewise for even. 
This partial decoupling
into an even and an odd sector
results from the parity
properties of the spherical harmonic functions $Y_\ell^m(\theta)$, which transform as $Y_\ell^m(\theta) \mapsto (-1)^{\ell+m} Y_\ell^m(\theta)$ under parity inversion---a divergence from the full spherical harmonic $Y_\ell^m(\theta,\phi) = Y_\ell^m(\theta) e^{i m \phi}$, which
satisfies
$Y_\ell^m(\theta, \phi) \mapsto (-1)^\ell Y_\ell^m(\theta,\phi)$.

Thus, the most natural way to label our two sectors are by the parity of $\ell+m$ to which they correspond. 
We may capture the two sectors by defining new matrices consisting of the elements of the matrix $M^{\ell \ell'}$ with the appropriate $\ell$ values: $\ell=2n+|m|$ for even $\ell+m$ and $\ell=2n + |m| + 1$ for odd $\ell+m$, with integer index $n$. 
Explicitly, we have
\begin{subequations}
\label{eq:angular_even_odd_matrix_eqs}
\begin{align}
    \label{eq:angular_even_odd_matrix_eqs_even}
    \sum_{n=0}^\infty M_E^{n n'} c_n ={}& 0\\     \label{eq:angular_even_odd_matrix_eqs_odd}
    \sum_{n=0}^\infty M_O^{n n'} d_n ={}& 0\, ,
\end{align}
\end{subequations}
where $c_n = b_{2n+|m|}$, $d_n = b_{2n+|m|+1}$, and 
\begin{subequations}
\begin{align}
    M_E^{n n'} \coloneq{}& M^{(2n+|m|), (2n' + |m|)}\qquad &&\text{(Even $\ell+m$)}\\
    M_O^{n n'} \coloneq{}& M^{(2n+|m|+1), (2n' + |m|+1)} &&\text{(Odd $\ell+m$)}\, ,
\end{align}
\end{subequations}
where $M^{\ell \ell'}$ is defined in Eq.~\eqref{eq:angular_matrix_def}
The matrices $M_E^{n n'}$ and $M_O^{n n'}$ encode the modes for which $\ell+m$ is even and odd, respectively. 
Each of these matrices is band diagonal, with entries on the first and second off-diagonals. 

We may now proceed in  several different directions. 
Eqs.~\eqref{eq:angular_even_odd_matrix_eqs} each admit nontrivial solutions only when their associated matrix has a nontrivial kernel, and therefore a vanishing determinant. 
Thus, if we cared only about finding valid values of the separation constant $\nu$, then we could simply seek solutions to the polynomial equations $\det\left(M_E^{n n'}\right) = 0$ and $\det\left(M_O^{n n'}\right) = 0$---in general a numerical procedure---as in Ref.~\cite{Dolan:2018dqv}. 
Here, however, we are primarily concerned with constructing explicit solutions for the angular function $S(\theta)$. 
In general, we could compute solutions using the nonlinear inverse iteration procedure employed in Ref.~\cite{Baumann:2019eav}, which iteratively solves for $b_n$ and $\nu$ simultaneously. 
However, we find that this procedure is unnecessary here, as exact solutions exist for the angular function in all our cases of interest. 


\subsection{Exact solutions for $\omega = \mu$}

We intend to model a \bh{} in equilibrium with a cold vector dark matter cloud, oscillating as $e^{-i \mu t}$.
We are thus interested in solutions for which $\omega = \mu$.
It was shown in Ref.~\cite{Dolan:2018dqv}
that the angular equation \eqref{eq:fkks_angular_eq_final} admits several exact solutions in the case $\omega = \mu$. 
We explore and expand on those here, extracting the ones most useful for the physical scenarios of interest.

Under the restriction $\omega = \mu$, both coefficient matrices, $M_E^{n n'}$ and $M_O^{n n'}$, become tridiagonal. This simplifies Eqs.~\eqref{eq:angular_even_odd_matrix_eqs} considerably, and it allows us to seek solutions of simpler form than in the general case. 
For the coefficients in Eqs.~\eqref{eq:angular_even_odd_matrix_eqs},
we thus take the ans\"{a}tze 
\begin{subequations}
\label{eq:ansatzCoeffscd}
\begin{align}
\label{eq:ansatzCoeffsc}
    \{c_n\} = \{1, c_1,0,0,\ldots\}
\,,\\
\label{eq:ansatzCoeffsd}
    \{d_n\} = \{1, d_1,0,0,\ldots\}
\end{align}
\end{subequations}
which amount to assuming an $S(\theta)$ of the form
\begin{equation}
    \label{eq:S_sol_general_special_form}
    S(\theta) = Y_\ell^m(\theta) + b_{\ell+2} Y_{\ell+2}^m(\theta)\, ,
\end{equation}
with the restrictions $m=\pm\ell$ for the even-($\ell+m$) sector and $m=\pm(\ell-1)$ for the odd-$(\ell+m)$ sector.
Our task is now to fix the separation constant $\nu$ and the remaining coefficient $b_{\ell+2}$. 


\subsubsection{$s=-1$, $m=\pm \ell$ solutions}

Beginning with the even modes, we consider the case $\ell = |m|$. Taking the additional restriction $c_1 \coloneq b_{\ell+2} = 0$
in the ansatz in Eq.~\eqref{eq:ansatzCoeffsc},
the matrix equation~\eqref{eq:angular_even_odd_matrix_eqs_even} reduces to two equations, $M_E^{00}=0$ and $M_E^{01}$=0. 
Both of them share a common pre-factor 
$\mu + \nu\left(\ell \mp a \mu\right)$ 
for the respective cases $m = \pm \ell$. 
Setting this pre-factor to zero, we find a set of solutions
for the separation constant
$\nu$, 
\begin{equation}
    \label{eq:sep_const_s=-1}
    \nu = \frac{-\mu}{\ell \mp a \mu}\, , \quad m=\pm \ell\, ,
\end{equation}
corresponding to an $S(\theta)$ solution
\begin{equation}
    \label{eq:S_func_s=-1}
    S(\theta) = Y_\ell^{\pm\ell}(\theta)\, .
\end{equation}
In the limit $a \to 0$, the monopole ($\ell = 0$) mode diverges, but the $\ell \geq 1$ modes remain finite, 
reducing to $\nu = -\mu / \ell$, the same as the $(s=-1)$ modes in Schwarzschild. 
We have thus found a set of vector-type modes in the even-parity sector. These are the $m = \pm \ell$
modes of the $s = -1$ polarization 
(for which $\ell \geq1$).

As shown in Refs.~\cite{Pani:2012vp, Pani:2012bp, Dolan:2018dqv}, superradiant 
quasi-bound states of the $s=-1$ polarization, with frequency $\omega = \omega_R + i \omega_I$, enjoy the shortest instability timescale, $\tau_\text{ins} =1/\omega_I$. This is due to their low total angular momentum $j = \ell + s$, compared to the other modes, which minimizes angular-momentum repulsion from the \bh{} and allows the field to maximally
accumulate 
in the \bh{'s} ergoregion. 
We expect this to also be reflected in our steady-state accretion rates. 

We have thus shown that in the $\omega=\mu$ case, the field's $(s,\ell,m) = (-1,1,1)$ mode, expected to be the dominant superradiant mode, is given by a pure $\ell = m = 1$ dipole, found from Eq.~\eqref{eq:S_func_s=-1}, with angular separation constant given by Eq. \eqref{eq:sep_const_s=-1}.


\subsubsection{$s=-1$, $m = \pm (\ell-1)$ and $s=0$, $m = \pm \ell$ solutions}

We now consider modes with odd $\ell + m$.\footnote{The careful reader may notice that the selection $\ell = \pm m$ makes $\ell+m$ strictly even. The $s=0$, $m = \pm \ell$ solutions are derived from the odd-parity matrix, but we make the re-indexing $\ell-1 \to \ell$ for physical reasons, a choice described in detail later in this section.}
We may extract a set of exact solutions from the odd-$(\ell+m)$ characteristic equation \eqref{eq:angular_even_odd_matrix_eqs_odd} by choosing $m = \pm(\ell - 1)$ and $d_1 = 0$
in the ansatz in Eq.~\eqref{eq:ansatzCoeffsd}. 
This yields two equations $M_O^{00} = 0$ and $M_O^{01} = 0$, from which we may extract the common polynomial equation
\begin{equation}
    \label{nu_charac_eq_lm1_odd}
    a \nu^2 + (a \mu \mp \ell) \mp \mu = 0\, , \quad m = \pm(\ell-1)\, ,
\end{equation}
which has solutions
\begin{subequations}
\label{eq:nu_lm1}
\begin{alignat}{2}
\label{eq:nu_lm1_plus}
\nu_+^{\pm} = {}& \frac{1}{2 a} \left( \ell - a \mu \pm \sqrt{(a \mu - \ell)^2 + 4 a \mu} \right)\,,\,\,\, {}&& m = (\ell-1) \\
\label{eq:nu_lm1_minus}
    \nu_-^{\pm} = {}& \frac{-1}{2 a} \left( \ell + a \mu \pm \sqrt{(a \mu + \ell)^2 - 4 a \mu} \right)\,,\,\,\, {}&& m = -(\ell - 1)\, .
\end{alignat}
\end{subequations}
We note that Eqs.~\eqref{nu_charac_eq_lm1_odd}-\eqref{eq:nu_lm1} differ from their corresponding expressions in Ref.~\cite{Dolan:2018dqv}, given in Eq.~(37) and the above paragraph of that work. We believe the expressions in this work to be correct after performing multiple checks and corresponding with the author of that work.\footnote{
In particular, the $\mp \mu$ term in Eq.~\eqref{nu_charac_eq_lm1_odd}
has a purely ``$-$" sign in Ref.~\cite{Dolan:2018dqv}, which carries forward into a sign difference between the $\nu$ solutions in Eq.~\eqref{eq:nu_lm1} of this work and those in Eq.~(37) of that work. 
We checked the solutions in Eqs.~\eqref{nu_charac_eq_lm1_odd}-\eqref{eq:nu_lm1} by plugging them into Eq.~\eqref{eq:fkks_angular_eq_final}. 
}
As before, to appropriately label these modes, we consider their behavior in the limit $a \to 0$, labeling them based on which Proca-field polarization they correspond in the Schwarzschild case. 

\paragraph*{Case 1 ($\nu_\pm^-$):}The solutions $\nu_+^-$ and $\nu_-^-$ each reduce to $-\mu / \ell$ in the limit $a \to 0$, meaning we may associate these with the even-parity vector-type $(s=-1)$ Schwarzschild modes with $m = \pm (\ell-1)$. 

\paragraph*{Case 2 ($\nu_\pm^+$):} 
Despite formally diverging in the limit $a\to0$, 
the solutions $\nu_-^+$ and $\nu_+^+$ do produce a physical solution,
as was shown in Ref.~\cite{Dolan:2018dqv}.
In particular, if one considers the radial equation~\eqref{eq:fkks_radial_eq_final} with $\nu = \nu_\pm^+$
and takes the $a \to 0$ limit carefully, then one recovers precisely the $s=0$ radial equation from Ref.~\cite{Rosa:2011my}, 
but with the identification $\ell-1 \to l$. 
This case thus corresponds to the $s=0$ polarization with $m =\pm l$, using the labeling from Ref.~\cite{Rosa:2011my}. 

In fact, we can demonstrate this correspondence even further. Since we are interested in co-rotating $(m>0)$ modes,
with $\ell=m+1$, 
we take the $\nu^+_+$ solution from Eq.~\eqref{eq:nu_lm1_plus}.
Using Eqs. (\ref{eq:fkks_ansatz}, \ref{eq:fkks_Z_ansatz}, \ref{eq:S_sol_general_special_form}),
the vector potential
then reads
\begin{equation}
    \label{eq:s=0_Amu_reconstruct}
    A^\mu = B^{\mu\nu}\nabla_\nu \left[R(r) Y_{m+1}^{m}(\theta) e^{i m \phi} e^{-i \mu t} \right]\, .
\end{equation}
This expands into a set of lengthy expressions, but all terms contain a factor of either $Y_{m+1}^m(\theta)$ or $Y_{m+1}^{m+1}(\theta)$. From the explicit expressions for the spherical harmonics, one can derive the useful ladder identities
\begin{subequations}
\begin{align}
    Y^{m+1}_{m+1}(\theta) ={}& -\sqrt{\frac{2m+3}{2m+2}} \sin\theta\, Y^m_m(\theta)\\
    Y_{m+1}^m(\theta) ={}& \sqrt{2m+3} \cos\theta\, Y_m^m(\theta)\, .
\end{align}
\end{subequations}
Using these, one can reduce Eq. \eqref{eq:s=0_Amu_reconstruct} to an expression of the form
\begin{equation}
    \label{eq:Amu_gen_mm_s=0_form}
    A_\mu = \mathcal{B}_\mu(r,\theta) Y_m^m(\theta, \phi) e^{-i \mu t}\, .
\end{equation}
The complicated coefficient $\mathcal{B}_\mu(r, \theta)$ 
can be expanded in $a \mu$, finding, at leading order,
\begin{align}
\mathcal{B}_\mu(r,\theta) = {} & 
    -\sqrt{2m + 3}\left(0, 0, \frac{1}{\sin\theta}, i \cos\theta \right) R(r) 
\nonumber \\ {} &
    + \mathcal{O}(a \mu)
\,.
\end{align}
For $\ell=m$ and $\omega=\mu$, the $s=0$ \vsh{} expression for the vector field $A_\mu$ is
\begin{align}
& \left(A_\mu^\text{VSH}\right)_{\ell=m}^m 
\\ & = 
    \frac{-i}{1+m}\left(0,0,\frac{1}{\sin\theta}, i\cos \theta \right) R(r) Y_m^m(\theta,\phi) e^{-i \mu t}
\nonumber\,.
\end{align}
We thus see that the 
vector field
expression from Eq.~\eqref{eq:s=0_Amu_reconstruct}, 
temporarily
labeled 
as
$\left(A_\mu^\text{FKKS}\right)_{\ell=m+1}^m$,
is given by
\begin{align}
\label{eq:fkks_to_vsh_s=0}
\left(A_\mu^\text{FKKS}\right)_{\ell=m+1}^m 
= {} &
    i (m+1) \sqrt{2 m+3} \left(A_\mu^\text{VSH}\right)_{\ell=m}^m 
\nonumber \\ {} &
    + \mathcal{O}(a \mu)
\,,
\end{align}
to leading order in $a \mu$.
Thus, 
for vanishing \bh{} spin,
$a \to 0$, 
the \lfkks{} vector field solution $\left(A_\mu^\text{FKKS}\right)_{\ell=m+1}^m$ is given, up to constant factors, precisely by 
the \vsh{} solution
$\left(A_\mu^\text{VSH}\right)_{\ell=m}^m$. 

We have now arrived at an unfortunate ambiguity.
The $s=0$ modes can either be labeled by the value $\ell=m+1$, 
corresponding to the spherical harmonic in Eq.~\eqref{eq:s=0_Amu_reconstruct}, 
or by $\ell = m$, as in Eq.~\eqref{eq:Amu_gen_mm_s=0_form}, 
corresponding to the \vsh{} mode that they reduce to in the $a \to 0$ limit, shown in Eq. \eqref{eq:fkks_to_vsh_s=0}. 
The former is more self-contained to this calculation, but it breaks a simple association of $\ell$ with the field's total angular momentum, usually given by $j = \ell+s$ \cite{Rosa:2011my, Hancock:2025ois}; the latter requires reference to the \vsh{} approach from other work, but it gives an $\ell$ which yields the correct total angular momentum and matches the labeling of the other polarizations. 
For these reasons, we will take the latter approach, labeling
the $\nu_+^+$ modes 
as $(s,\ell,m) = (0,m,m)$, as identified by their behavior in the $a \to 0$ limit. 

At this point, we have found solutions for the $(s,\ell,m)=(-1,1,1)$ and $(0,1,1)$ modes. We now procede to the $(1,0,0)$ and $(1,1,1)$ modes.


\subsubsection{$s=1$, $m=\pm \ell$ solutions}

We finally return to the even-$(\ell+m)$ equation in Eq.~\eqref{eq:angular_even_odd_matrix_eqs_even}, this time keeping the second term in Eq. \eqref{eq:S_sol_general_special_form}.
For $c_1 \neq 0$ 
in the ansatz in Eq.~\eqref{eq:ansatzCoeffsc},
the even-parity equation \eqref{eq:angular_even_odd_matrix_eqs_even} reduces to three polynomial equations
\begin{subequations}\label{eq:s=1_charac_eqs}
\begin{align}
    \label{eq:s=1_charac_eqs_1}
    0 ={}& M_E^{00} + c_1 M_E^{01} \\
    \label{eq:s=1_charac_eqs_2}
    0 ={}& M_E^{10} + c_1 M_E^{11}\\
    \label{eq:s=1_charac_eqs_3}
    0 ={}& M_E^{21}\, .
\end{align}
\end{subequations}
The third equation \eqref{eq:s=1_charac_eqs_3} is a cubic polynomial in $\nu$, containing $m$ as a free parameter,
while Eqs.~(\ref{eq:s=1_charac_eqs_1}--\ref{eq:s=1_charac_eqs_2}) are more complicated polynomials in $\nu$.

We begin by considering the $m = \ell =1$ case, in which Eq.~\eqref{eq:s=1_charac_eqs_3} reads
\begin{equation}
    \label{eq:l=m=1_nu_charac_eq}
    \left(1 - a\mu \right) a\, \nu^3 + \left(2 a \mu - a^2 \mu ^2 - 6\right)\nu ^2 +\mu\, \nu + \mu ^2 = 0\, .
\end{equation}
All three solutions for the separation constant $\nu$ 
give formal solutions to Eq.~\eqref{eq:fkks_angular_eq_final}. 
However, in the limit $a \to 0$, the largest and smallest 
roots
have behavior which correspond to no known Schwarzschild modes. 
Conversely, the middle root approaches $\mu / (\ell + 1)$ as $a \to 0$
as expected of the $s=1$ mode. 
We thus take the middle root to be the valid $s = 1$ solution for $m = \ell = 1$. 

With our solution for the separation constant $\nu$ in hand, the undetermined coefficient $c_1$ can be found by a simple algebraic inversion of either Eq.~\eqref{eq:s=1_charac_eqs_1} or Eq.~\eqref{eq:s=1_charac_eqs_2}, which yield the same result.
The result is a function depending only on the dimensionless parameter $a \mu$, $c_1 = c_1(a \mu)$. 
From the $\nu$ and $c_1$ solutions, we may reconstruct the full $S(\theta)$ solution using Eq.~\eqref{eq:S_sol_general_special_form}. For $\ell=1$, the solution is a dipole plus an $\mathcal{O}\left((a \mu)^2\right)$ correction. Explicitly, we have 
\begin{align}
\label{eq:s=l=1_angular_sol}
S(\theta) & = 
    \sin \theta \left(1 + c_1(a\mu) \frac{1}{2}\sqrt{\frac{7}{2}} \left(5 \cos^2\theta - 1\right)\right)
\\ & \approx 
    -2\sqrt{\frac{2\pi}{3}}\left(Y_1^1(\theta) + \left[\frac{3 (a \mu)^2}{25\sqrt{14}} + \mathcal{O}\left((a\mu)^3\right)\right] Y_3^1(\theta)\right)
\,.\nonumber
\end{align}
In practice, we will simply use the exact solution, but the approximation provides useful insight into its behavior.

We next consider
$m=\ell=0$, for which the characteristic polynomial reads
\begin{equation}
    \label{eq:l=m=0_nu_charac_eq}
    a^2 \mu\, \nu^3 + (2 + a^2\mu^2)\nu^2 - \mu\, \nu - \mu^2 = 0\, .
\end{equation}
Again, all three roots of Eq.~\eqref{eq:l=m=0_nu_charac_eq}
are valid solutions.
Here, only the largest root reduces to a  Schwarzschild solution,
approaching $\nu \to \mu$ as $a \to 0$, which is expected for the 
$(s, \ell, m)=(1,0,0)$ mode.
The other roots have no correspondence to (known) Schwarzschild solutions in the $a\to0$ limit.
We thus select 
the largest root of Eq.~\eqref{eq:l=m=0_nu_charac_eq}
as the physical root.
We may gain additional insight by considering the behavior of $c_1$, which can be fixed using either Eq. \eqref{eq:s=1_charac_eqs_1} or Eq. \eqref{eq:s=1_charac_eqs_2}.
For positive \bh{} spin, $c_1$ is nonzero and positive, but it smoothly approaches zero as $a \to 0$. Our
$m=\ell=0$ solution is thus exactly the Kerr counterpart of the Schwarzschild $s=1$ monopole mode. Explicitly, it is a monopole with an $\mathcal{O}((a\mu)^2)$ quadrupolar correction, given by
\begin{align}
\label{eq:S_sol_l=m=0}
S(\theta) = {}& 1 + c_1(a\mu)\frac{\sqrt{5}}{2} \left(3 \cos^2 \theta - 1\right)
\\ \approx {}& 
    2\sqrt{\pi}\left(Y_0^0(\theta) -  \left[\frac{2(a \mu)^2}{9 \sqrt{5}} + \mathcal{O}\left((a \mu)^4\right) \right]Y_2^0(\theta) \right)
\,.\nonumber
\end{align}
As the lowest angular momentum mode, we expect this mode to be the dominant mass accretion channel in astrophysical contexts. 

\subsection{Summary of relevant modes}

For our study, we are interested principally in 
\begin{enumerate}
    \item The monopole mode, $(s,\ell,m) = (1,0,0)$, which we expect to be the dominant contribution for astrophysical wave dark matter accretion
    \item The fundamental co-rotating mode for each polarization, $(s,\ell,m) = (s,1,1)$, which are the dominant superradiant modes.
\end{enumerate}

We have now derived analytic expressions for the separation constant $\nu$ and the angular function $S(\theta)$ in each of these cases, under the restriction $\omega=\mu$. For the reader's convenience, these are collected in Table~\ref{tab:modes_summary}.

\begin{table}[!ht]
    \centering
    \caption{Summary of angular equation solutions}
    \label{tab:modes_summary}
    \begin{tabular}{c|l @{\hspace{2em}} l}
        \toprule
        $(s, \ell, m)$&   $\nu$&$S(\theta)$\\
        \midrule
        $(1,0,0)$ & Eq. \eqref{eq:l=m=0_nu_charac_eq} (greatest root) & Eq. \eqref{eq:S_sol_l=m=0} \\
        $(1,1,1)$& Eq. \eqref{eq:l=m=1_nu_charac_eq} (middle root) & Eq. \eqref{eq:s=l=1_angular_sol} \\
        $(0,1,1)$& Eq. \eqref{eq:nu_lm1_plus} $(\nu^+_+)$ & $Y_2^1(\theta)$\\
        $(-1,1,1)$& Eq. \eqref{eq:sep_const_s=-1} (``$+$" choice) & $Y_1^1(\theta)$\\
        \bottomrule
    \end{tabular}
\end{table}

\section{Solving the radial equation}
\label{sec:radial_eq_solving}

With the angular equation~\eqref{eq:fkks_angular_eq_final} solved, we may now solve the radial equation~\eqref{eq:fkks_radial_eq_final} for suitable values of the separation constant $\nu$. 
This may be done in much the same manner as in Paper I, a method we summarize in this section.
For both numerical stability and physical insight, it is convenient to work in the tortoise-like radial coordinate $r_*$, implicitly defined by $\dif r_* = \frac{r^2 + a^2}{\Delta} \dif r$.
As shown explicitly in Appendix \ref{sec:radial_eq_asymptotics}, in the near-horizon limit $r_* \to - \infty$, 
Eq. \eqref{eq:fkks_radial_eq_final} has asymptotic solutions
\begin{equation}
\label{eq:RadialFctNearHorizon}
    R(r_*) = C_1 e^{i \left(\omega - m \Omega_H\right) r_*} + C_2 e^{-i \left(\omega - m \Omega_H\right) r_*}\, ,
\end{equation}
which are the expected outgoing and ingoing waves, respectively. 
Solutions with regular near-horizon behavior, something we wish to enforce, are those which reduce to only the ingoing solution in this regime.

The radial equation \eqref{eq:fkks_radial_eq_final} has a complicated singularity structure and does not admit solutions in terms of confluent Heun or other standard special functions.
However, it is amenable to finite-difference numerical integration.
Thus, as in Paper I, we solve the radial equation~\eqref{eq:fkks_radial_eq_final} 
using \textsc{Mathematica}'s ``NDSolve" package.
As initial conditions for the solver, we take the value and first derivative 
of the expression
\begin{equation}
    R^{\text{Near}}(r_*) = e^{-i \left(\omega - m \Omega_H\right) r_*}\, ,
\end{equation}
representing a purely ingoing wave,
at a point in the region
$r_* \lesssim 0$, where the near-horizon approximation in Eq.~\eqref{eq:RadialFctNearHorizon} is
valid. This produces numerical solutions which are purely ingoing and normalized to $|R|=1$ near the horizon.

To understand the behavior of the radial function far from the horizon, we note that the Kerr metric in Eq.~\eqref{eq:kerr_metric} can be expanded as 
$g_{\mu\nu} = g_{\mu\nu}^\text{Sch.} + \frac{a}{r} g_{\mu\nu}^{(1)} + \left(\frac{a}{r}\right)^2 g_{\mu\nu}^{(2)} + \ldots$,
where $g_{\mu\nu}^\text{Sch.}$ is the Schwarzschild metric. 
Spin contributions thus enter at $\mathcal{O}(a/r)$, so for $r \gg a$, Schwarzschild results become approximately valid. 
Far from the horizon, solutions to the Kerr radial equation \eqref{eq:fkks_radial_eq_final} thus have the same behavior as their Schwarzschild counterparts~\cite{Hancock:2025ois},
\begin{equation}
\label{eq:odd-par_large_r_approx}
    R(r) \approx C_{\text{in}}(\mu M)r^{1/4}e^{-2 i \mu \sqrt{2M r}} 
    + C_{\text{out}}(\mu M) r^{1/4} e^{2i\mu \sqrt{2M r}}
\,.
\end{equation}
As in the Schwarzschild case, there is no exact functional form for the coefficients $C_\text{in/out}=C_\text{in/out}(\mu M)$,
but they may be extracted from the behavior of the numerical solutions for $R(r)$ in the regime $r \gg 1/(\mu^2 M)$.
Here, we focus on the behavior of astrophysically relevant quantities derived from the field's energy-momentum tensor.

\section{Results}
\label{sec:results}

With the solutions for 
the radial and angular functions
$R(r)$ and $S(\theta)$ in hand, we may reconstruct full 
vector-field
solutions to the Proca equation~\eqref{eq:ProcaEoM} using the ansatz in Eq.~\eqref{eq:fkks_ansatz}. 
In this section, we use these solutions,
along with the field's energy-momentum tensor given in Eq.~\eqref{eq:ProcaTmunu},
to calculate the astrophysical quantities of interest---namely, the field's density profile and the rates at which the \bh{} accretes or loses
mass and angular momentum.
Note that while we generally characterize the \bh{} spin using its spin parameter $a$ in previous sections, here, we instead use its dimensionless spin $\chi = a/M$.

We seek a density profile which matches the density of the dark matter environment at the edge of the \bh{'s} sphere of influence. 
The radius of this sphere, $r_c$, is roughly the distance at which the \bh{} gravitational potential rivals that of the surrounding matter.
Standard virial theorem arguments prompt us to use $r_c \sim 10^6 M$~\cite{Hui:2019aqm, Ghez:1998ph, Hancock:2025ois}.
As such, all results in this section are normalized such that the field's density at a distance $r=r_c$ is $\rho = \rho_c$, where $\rho_c$ is an adjustable parameter for the ambient dark matter density and $r_c$ is the edge of the \bh{} sphere of influence. 
While, in principal, there are many dark matter densities we could pick for $\rho_c$, 
we select $\rho_c \sim 10\, \msun/\pc^3 \sim 100\, \Gev/\cm^3$.
This choice corresponds to
the core of a wave dark matter soliton at the center of a large galaxy like M87~\cite{Schive:2014hza}.
While much higher than dark matter densities in most of the universe,
this choice
represents a scenario in which the effects we consider are most likely to be observationally relevant. 

\subsection{Density profile}

We first consider the density profile formed by the Proca field around the \bh{}. 
In Paper~I, we found that all modes of the field form a density ``spike" around a Schwarzschild \bh{} if sourced from a dark matter bath.
In the particle regime ($\mu M\gtrsim 1$), 
the spike had a profile $\rho \sim r^{-3/2}$, which matches both the analogous scalar-field profile~\cite{Hui:2019aqm, Clough:2019jpm} and the density profiles of generic non-relativistic particle dark matter spikes studied in the literature~\cite{Bertschinger:1985pd, Gondolo:1999ef}. 
In the wave regime ($\mu M < 1$), 
the field formed standing waves, and the density profile exhibited notches at the standing-wave nodes.
Averaging over these radial variations, however, we found a smoothed profile which retained the shape $\rho \sim r^{-3/2}$. 
In this section, we extend these results 
from Paper~I
to the Kerr case, examining how 
the features described above
are changed 
when the 
\bh{} is rotating.

The field's density is given by
\begin{equation}
    \rho = u^\mu u^\nu T_{\mu\nu}\, ,
\end{equation}
where $u^\mu\partial_\mu \sim \partial_t$ is the (normalized) 4-velocity of a stationary observer and $T_{\mu\nu}$ is the energy-momentum tensor from Eq.~\eqref{eq:ProcaTmunu}. 
For this and all following results, we normalize $T_{\mu\nu}$ such that 
the spatially-averaged density 
evaluated at the edge of the \bh{'s} sphere of influence satisfies
\begin{equation}
    \label{eq:rho_avg_condition}
    \langle\rho\rangle|_{r=r_c} = \rho_c\, ,
\end{equation}
where
the spatially-averaged density is defined as
\begin{equation}
    \langle\rho\rangle|_r = \frac{\oint_{S^2} \dif \Omega \int r'^2\dif r' \rho(r',\theta,\phi) q(r,r')}{\oint_{S^2} \dif \Omega \int r'^2\dif r' q(r,r')}\, ,
\end{equation}
for a suitable smearing function $q(r,r')$, as in Paper I. 
In practice, for the smearing function, we use a standard boxcar or double Heaviside function centered at a radius $r$ with a sufficiently large width to contain a few radial oscillations.
When imposing Eq.~\eqref{eq:rho_avg_condition}, to match the values in Paper I, we use $r_c = 2\times10^6 M$. 
However, we emphasize that this is a strictly order-of-magnitude value, so all numerical results are only precise up to an order of magnitude. 

\begin{figure*}[t]
    \centering
    \includegraphics[width=0.45\linewidth]{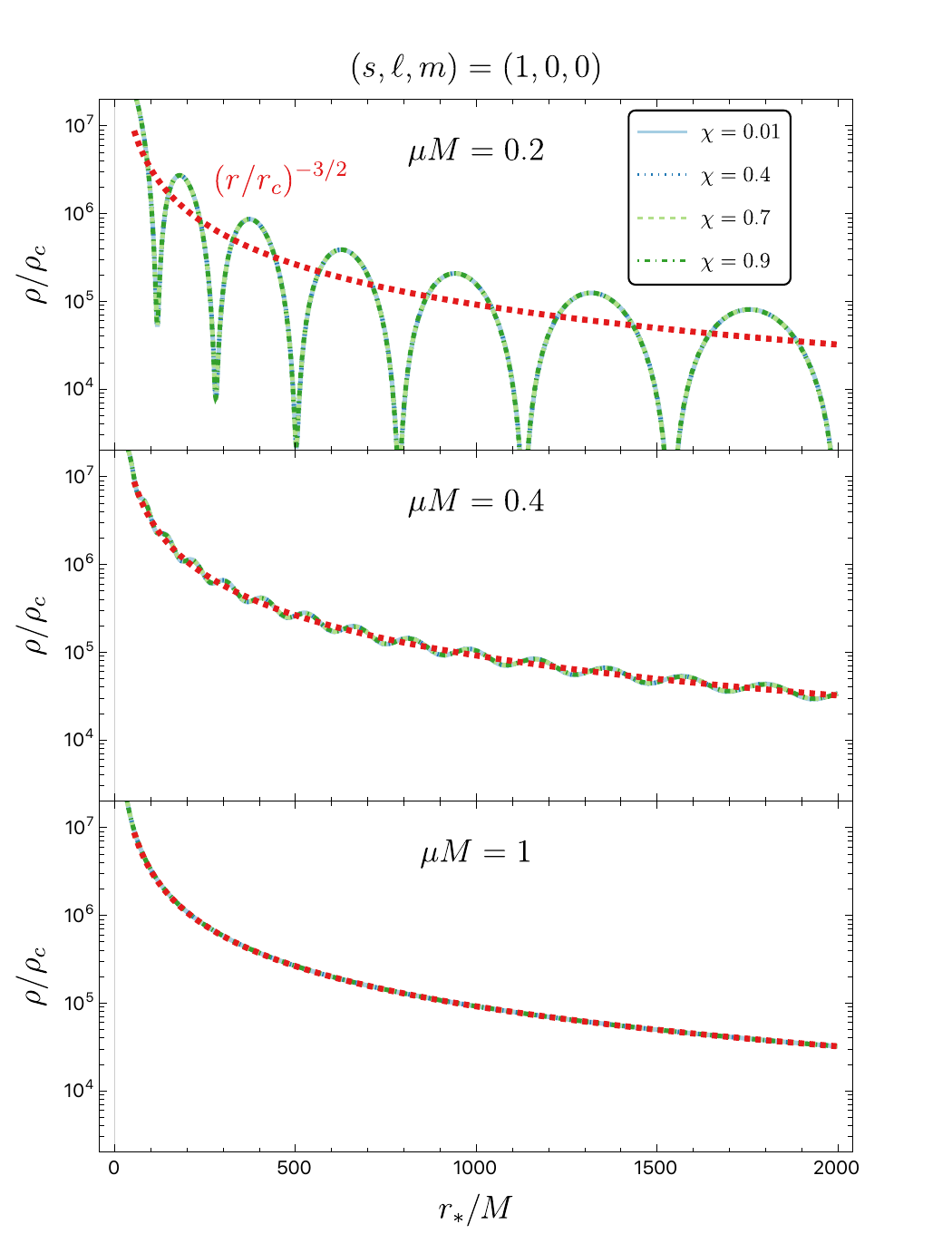}
    \includegraphics[width=0.45\linewidth]{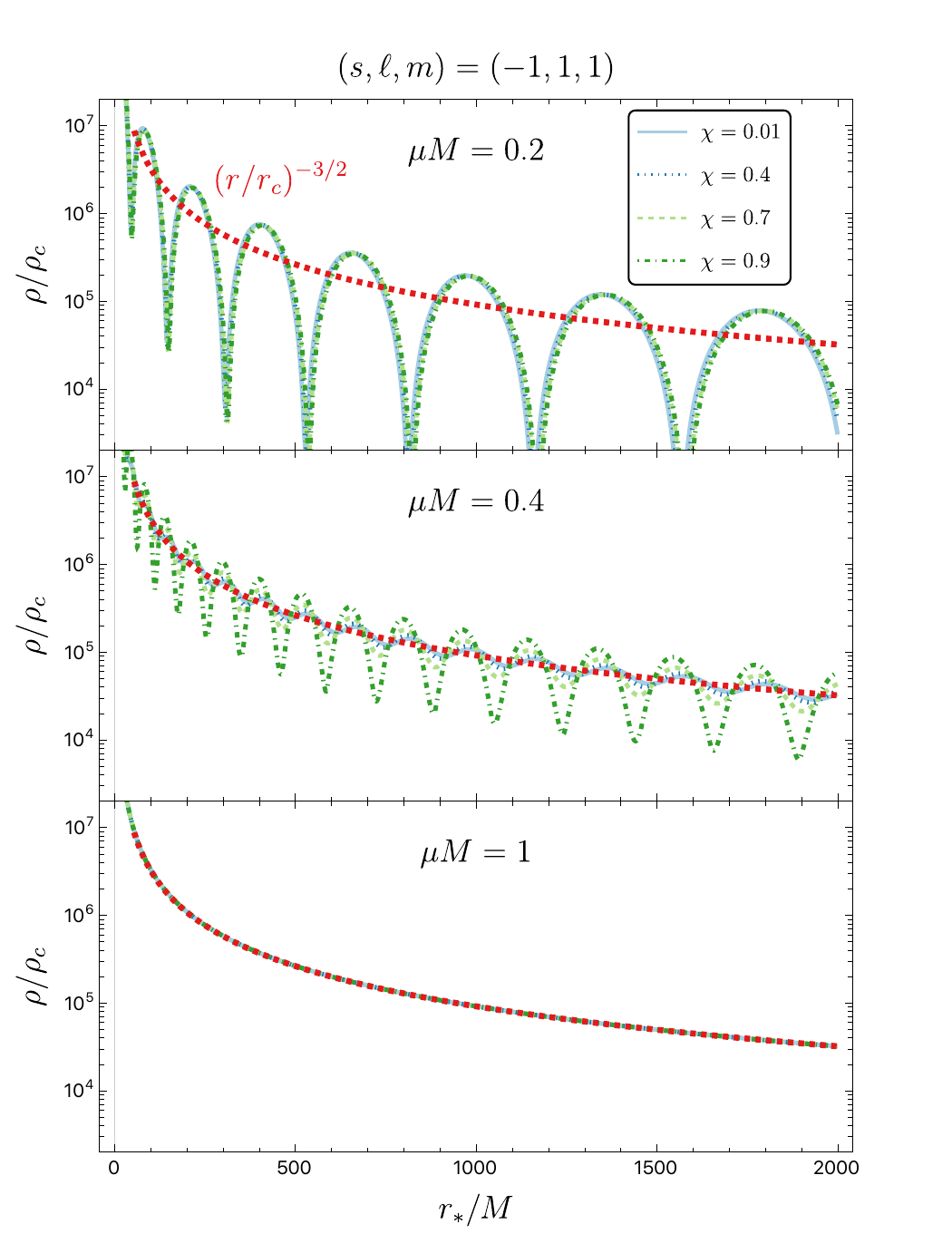}
    \includegraphics[width=0.45\linewidth]{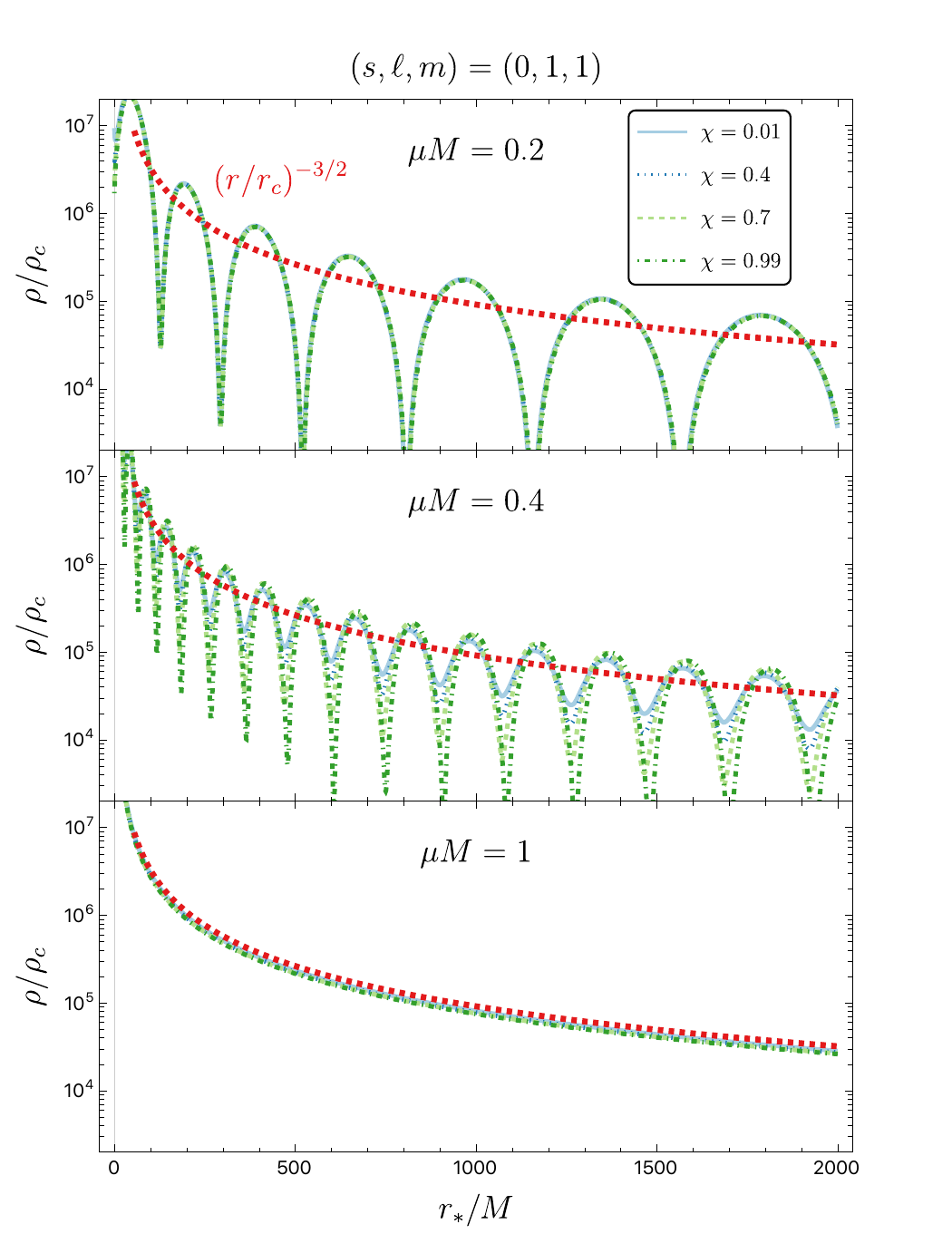}
    \includegraphics[width=0.45\linewidth]{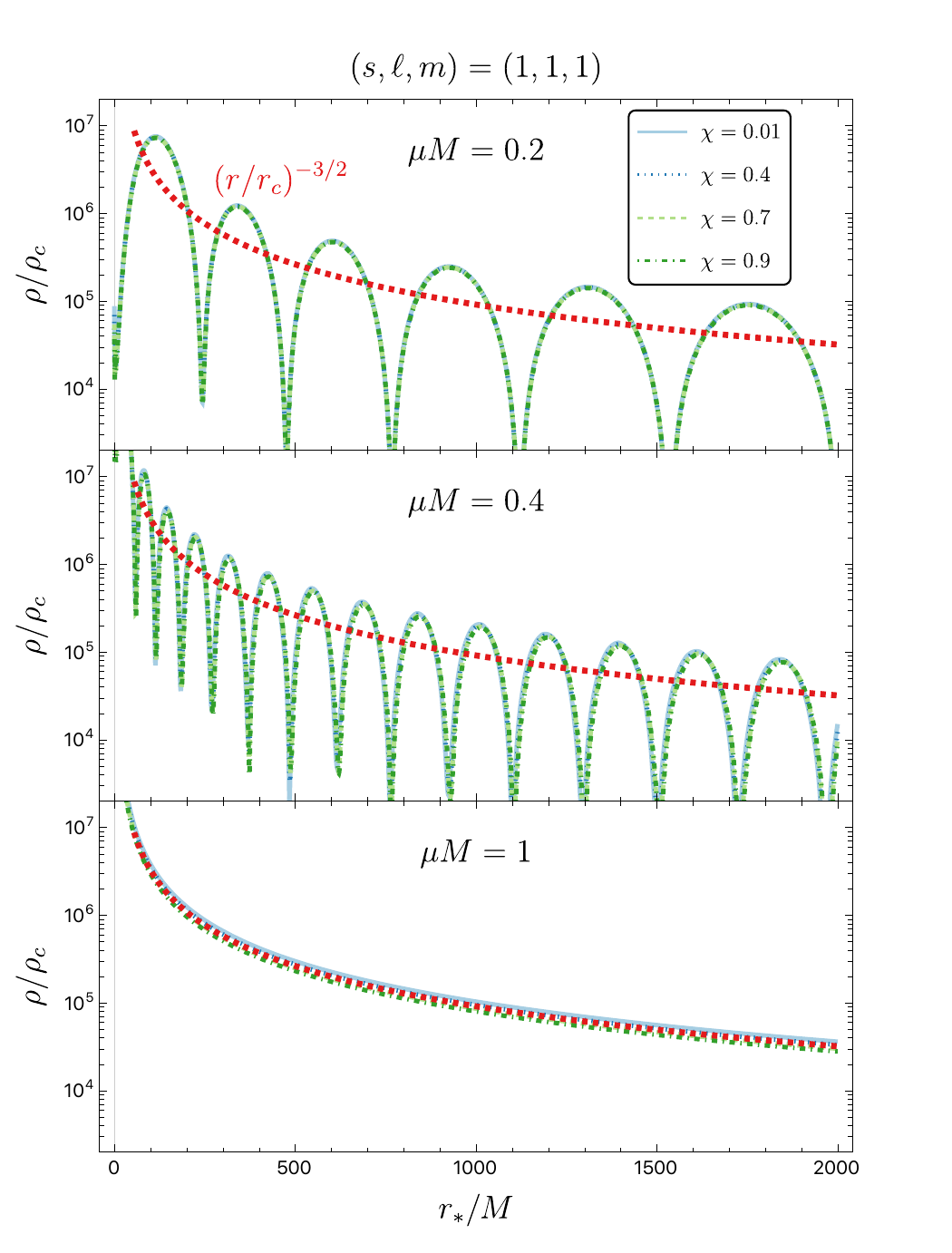}
    \caption{Density profiles 
    as a function of the tortoise-like radial coordinate 
    for the monopole $(s, \ell, m) = (1,0,0)$ mode (upper left) and for each polarization's lowest $m=1$ mode---$(-1,1,1)$ (upper right), $(0,1,1)$ (lower left), and $(1,1,1)$ (lower right).
    These profiles are all computed along $\theta=\pi/2$, except the $(0,1,1)$ mode, which is computed along $\theta=\pi/6$.
    The field's averaged density profile $\langle \rho \rangle|_r = \rho_c \left(r/r_c\right)^{-3/2}$ is superimposed as a red dotted line. 
    The three panels of each subplot show,
    from top to bottom,
    the density profile for
    dimensionless mass parameter values
    $\mu M \in \{0.2, 0.4, 1\}$ 
    at a range of \bh{} spins
    $\chi$.
    We see that the spin-induced distortion of the density profile is, in all cases, minor. The main effect is to shift the threshold between the the wave and particle regimes, leading to variation between the profiles at $\mu M=0.4$ for all modes.}
    \label{fig:density_plots}
\end{figure*}

\begin{figure}[th]
    \centering
    \includegraphics[height=1.5in]{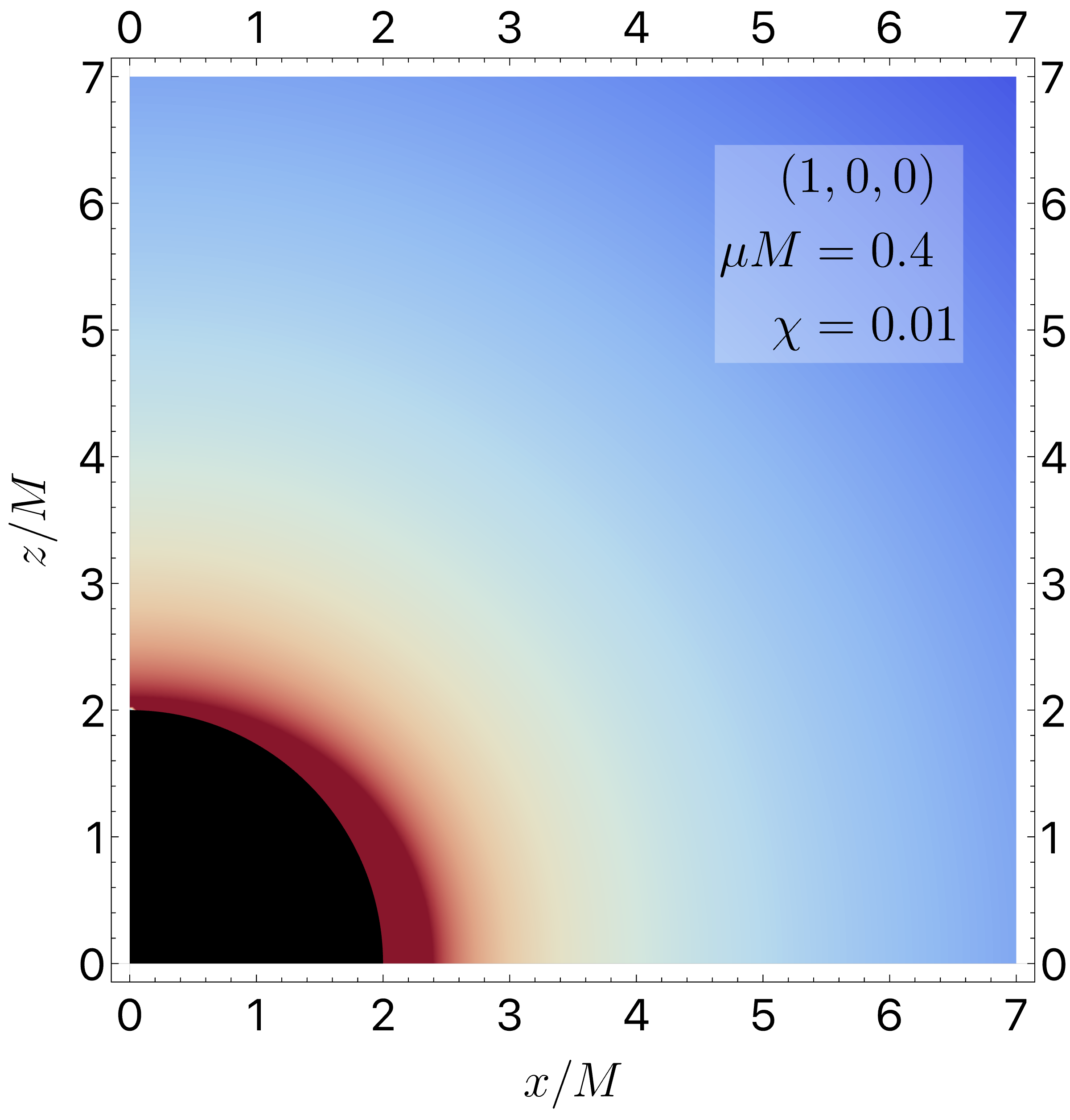}
    \includegraphics[height=1.5in]{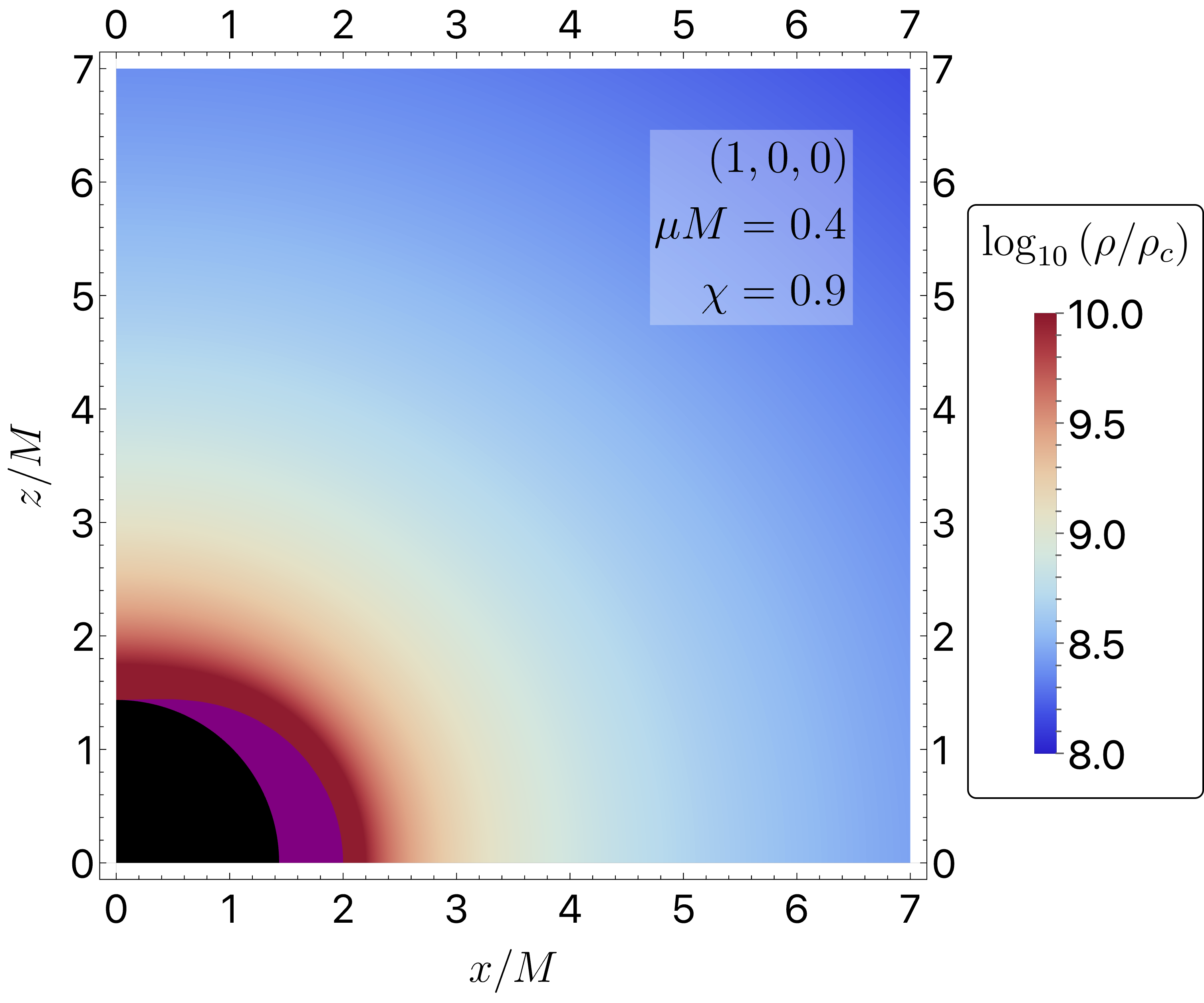}
    \includegraphics[height=1.45in]{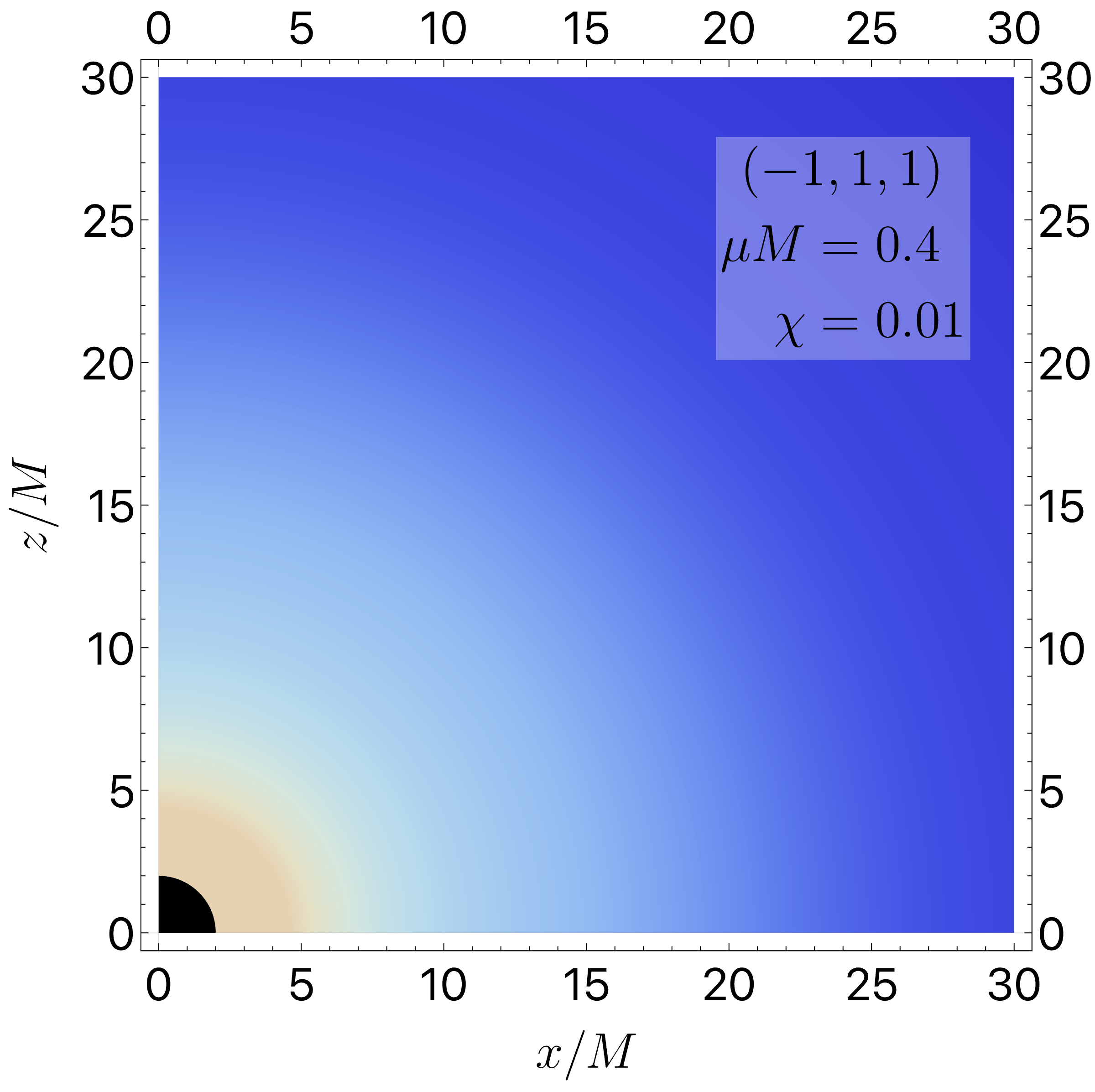}
    \includegraphics[height=1.45in]{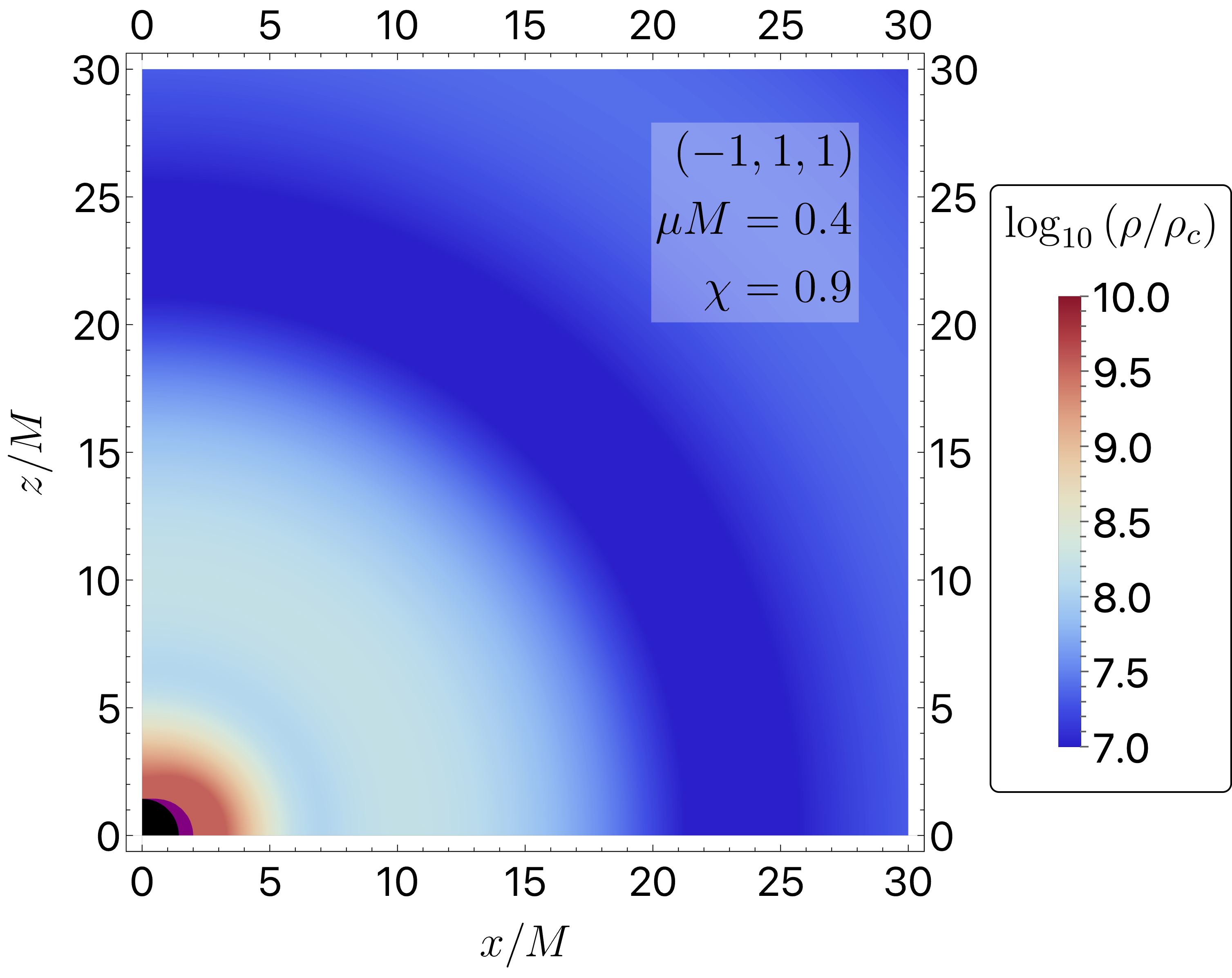}
    \caption{Plots of the field's density for $\mu M = 0.4$ across a quadrant of the $(x,z)$ plane. 
    The \bh{} horizon is shown in black, and its ergoregion
    is shown in purple. 
    The upper panels show the density of the 
    $(s,\ell,m) = (1,0,0)$ mode for 
    $\chi \in \{0.01, 0.9\}$, and the lower panels show the same for $(s, \ell, m) = (-1,1,1)$.
    The former mode has a monopole profile, which is only slightly distorted at high \bh{} spin. 
    The latter has a near-monopole profile, despite its nonzero angular momentum. At $\chi=0.9$, its central density spike is both more oblate and slightly larger than in the $(1,0,0)$ case, and underdense rings from the field's standing-wave pattern are visible. The plots in each row share the same color coding, shown in the colorbars on the right. }
    \label{fig:density_plots_2d}
\end{figure}

In Figure~\ref{fig:density_plots},
we plot the Proca field's density as a function of the tortoise-like coordinate, $r_{\ast}$, for the $(s,\ell,m) = (1,0,0)$ mode and for the $(s,1,1)$ modes 
for \bh{} spins $\chi \in \{0.01,0.4,0.7,0.9\}$.
Here, the density
is evaluated 
along the $\theta=\pi/2$ line
for all modes except $(0,1,1)$, which we evaluate at $\theta=\pi/6$ since it has a density node at the former value. 
The dependence of the density on the angle $\theta$ is shown in Figure~\ref{fig:density_plots_2d}
which we will discuss below.
For all modes, the density has no dependence on $t$ or $\phi$.
As in the scalar-field case, \bh{} rotation does little to distort the field's equilibrium density profile, except very near the horizon~\cite{Bamber:2020bpu}.

Like in the Schwarzschild case, 
all vector-field modes have a smooth
density profile with
$\rho \sim r^{-3/2}$ 
in the particle  regime, $\mu M \gtrsim 1$; 
see the lower panels in Figure~\ref{fig:density_plots}.
In the wave regime, $\mu M < 1$,
all modes exhibit a bumpy density profile with standing-wave nodes producing the wave-regime notches;
see the upper and middle panels 
in Figure~\ref{fig:density_plots}.
Deep in both regimes, the effect of the \bh{} spin is almost un-noticeable in Figure~\ref{fig:density_plots}, even for high \bh{} spins of $\chi = 0.9$. 
In the intermediate  regime, $0.1 \lesssim \mu M < 1$, 
the notches deepen as the \bh{}
spin is increased for fixed mass parameter.

This happens because for 
non vanishing spins,
$\chi >0$, the presence of a superradiant regime at low field mass
(cf.~Eq.~\eqref{eq:superrad_condition} with $\omega=\mu$)---necessarily a wave-regime phenomenon---pushes the wave-particle cutoff to a larger mass parameter than for $\chi=0$. 
This effect is demonstrated in more detail in the next section. 
In all cases considered, 
the field forms an ``overdensity'' and reaches
$\rho \sim 10^7 \rho_c$ within $100 M$ of the horizon.

The effect of the \bh{} spin is clearer in the immediate vicinity of the horizon, as 
shown in Figure~\ref{fig:density_plots_2d}
where we display the Proca-field density
for a segment of the $(x,z)$ plane, defined by $(x,z) = (r\sin \theta, r\cos\theta)$.
The \bh{} horizon is shown in black, and its ergoregion is shown in purple.
Focusing on the choice $\mu M=0.4$ for the dimensionless mass parameter,
we consider the $(s,\ell,m)=(1,0,0)$ (upper panels) and 
$(-1,1,1)$ (lower panels) modes
on a Kerr background with spins $\chi=0.01$ (left panels) and $\chi=0.9$.
Note that both the spacetime and 
the field's
density configurations are cylindrically symmetric, so $\phi$ may be chosen arbitrarily. 
The
density profiles are also symmetric under $z \to -z$, so we need only consider one quadrant of the $(x,z)$ plane. 

The $(s,\ell,m) = (1,0,0)$ mode (upper panels) has the expected monopole shape for low \bh{} spin; 
see the upper-left panel of Figure~\ref{fig:density_plots_2d}.
For 
a high spin of
$\chi = 0.9$, this profile
becomes
slightly oblate within $\sim 2 M$ of the horizon, but is unaffected otherwise. 

At low \bh{} spin, the $(s,\ell,m)=(-1,1,1)$ mode 
exhibits a near-monopole profile, with no toroidal contribution visible;
see the lower-left panel of Figure~\ref{fig:density_plots_2d}.
This is reasonable because the expression for $\rho$, derived from Eq.~\eqref{eq:ProcaTmunu},
includes not only the $Y_1^1$ spherical harmonic, but also its derivatives, which do not vanish at the poles. These terms are of similar magnitude, and they conspire to create a density profile which is nearly monopole. 
For a high \bh{} spin of
$\chi=0.9$, this
density configuration is distorted; 
see the lower-right panel of Figure~\ref{fig:density_plots_2d}.
In particular,
the central density spike is sharpened and 
it becomes
oblate, while rings of relative underdensity are visible at $r \approx 7 M$ and $r \approx 24 M$. 
These rings correspond 
to
nodes in the field's standing-wave profile.
The ring at $r \approx 24 M$ coincides with the innermost density node visible in the $\mu M=0.4$ plot in the upper right panel of Figure~\ref{fig:density_plots}.

For all modes considered, the transition between the low-spin and high-spin morphologies is smooth, and no obvious change occurs at the superradiance threshold.
In general, we see that in each of these cases, the density exceeds $\rho \sim 10^9 \rho_c$ within $10 M$ of the horizon. 

Overall, we see that \bh{} spin does little to affect the field's equilibrium density profile anywhere except in the horizon's immediate vicinity
($r \lesssim 10 M$). 
Even with \bh{} rotation, therefore, the equilibrium configuration is a dark matter spike 
with a density profile $\rho\sim (r/r_{c})^{-3/2}$,
similar to that found in Paper~I.


We may, therefore, speculate that the 
dynamics/evolution/phenomenology
of a second \bh{} orbiting the primary \bh{} within a dark matter spike would,
to leading order,
remain unchanged by introducing rotation of the primary \bh{.} 
We note, however, that this does not rule out the possibility of the cloud dissipating or being otherwise disrupted before equilibrating by other dynamical effects, such as superradiance, which is relevant when $\Omega_H \gtrsim \mu$; see Eq. \eqref{eq:superrad_condition}. 
We investigate this more closely in later sections.


As in the Schwarzschild case, the field forms a ``spike" of enhanced density around the \bh{}. This can be understood as the field simply filling the \bh{'s} gravitational potential well, as expected from Newtonian physics. 
This is not significantly disrupted for a rotating \bh{}, except very near the horizon, as can be seen in Figure~\ref{fig:density_plots_2d}. As in the Schwarzschild case, the density reaches $\rho \sim 10^7 \rho_c$ within $100 M$ of the horizon and reaches $\rho \sim 10^9 \rho_c$ within $10 M$ of the horizon. 
Two quantities which are affected far more by \bh{} rotation are the rates of mass and angular momentum accretion. We turn to those next.


\subsection{Mass accretion}


The principal novel effect that \bh{} rotation introduces is the appearance of a superradiant regime, present for angular modes with $m > 0$.
Modes satisfying the superradiance condition 
in Eq.~\eqref{eq:superrad_condition} are amplified through interaction with the \bh{}, rather than attenuated. 
Standard perturbative superradiance studies~\cite{Detweiler:1980uk, Dolan:2007mj, Pani:2012bp, Dolan:2018dqv} search for quasi-bound states of the field,
solving for
a complex frequency
$\omega$ as an eigenvalue problem. 
Those states for which 
the real part of the frequency satisfies the superadiance condition,
$\omega_R < m \Omega_H$,
with $m>0$,
necessarily have 
a positive imaginary part of the frequency,
$\omega_I >0$,
indicating an exponentially growing field; this is the superradiant instability.
Here, we set 
a real frequency,
$\omega = \mu \in \mathbb{R}$, allowed because we leave our large-$r$ boundary condition free, so $\omega_I = 0$ and we see no superradiant instability. 
Nonetheless,
we find that
waves scattering off the \bh{} which satisfy the superradiance condition are indeed amplified, but our approach assumes that this occurs in equilibrium with an environment which carries off the excess energy.
To see the signature of this, we need just consider the rates at which the \bh{} accretes or loses mass and angular momentum due to its interactions with the field at relevant values of $(\chi,\mu)$. We begin with the mass accretion rate.

The \bh{} accretes mass from the field at a rate
\begin{equation}
    \label{eq:mass_accretion_rate_general_formula}
    \dot{M} = \oint_{S^2} \dif\Omega \left(r^2 + a^2 \cos^2\theta\right) \left \langle T^r{}_{t} \right \rangle |_{r}\, ,
\end{equation}
derived in Appendix~\ref{sec:acc_rates_defs}, where $T^r{}_t$, 
a component of the field's energy-momentum tensor,
defined in Eq.~\eqref{eq:ProcaTmunu},
may be evaluated at any $r$ outside the horizon.
We expect superradiance to manifest 
as a change in the sign of
$\dot{M}$;
it is uniformly positive for $\chi=0$ 
and in the accretion regime,
and it falls below zero, $\dot{M} < 0$, in the superradiant regime,
showing a change from \bh{} mass ``accretion" to mass ``extraction" by the field. 

The lowest modes with $m>0$ are the $(\ell,m) = (1,1)$ modes of each polarization, $s \in \{-1, 0, 1\}$.
With total angular momentum $j=\ell + s = 0$, the $s=-1$ polarization experiences the smallest angular momentum barrier of the three, allowing the field to 
accumulate in the \bh{} ergoregion, and 
thus is subject to the strongest superradiant amplification. 
We therefore expect 
the $s=-1$
polarization to extract the most \bh{} mass
in the superradiant regime, $\omega = \mu < m \Omega_H$. 
This hypothesis is supported by two facts: first, that 
quasi-bound states for this polarization experience the strongest superradiance instability~\cite{Dolan:2007mj, Pani:2012bp}, and 
second, that this polarization experiences a wave-regime mass accretion rate two to three order of magnitude larger than the others for $\chi = 0$~\cite{Hancock:2025ois}. 
Conversely, we expect the monopole mode,
$(s,\ell,m) = (1,0,0)$,
to be the dominant particle-regime accretion pathway. 
However, having $m=0$, it should experience no superradiance.

\begin{figure}[!ht]
    \centering
    \includegraphics[width=0.9\linewidth]{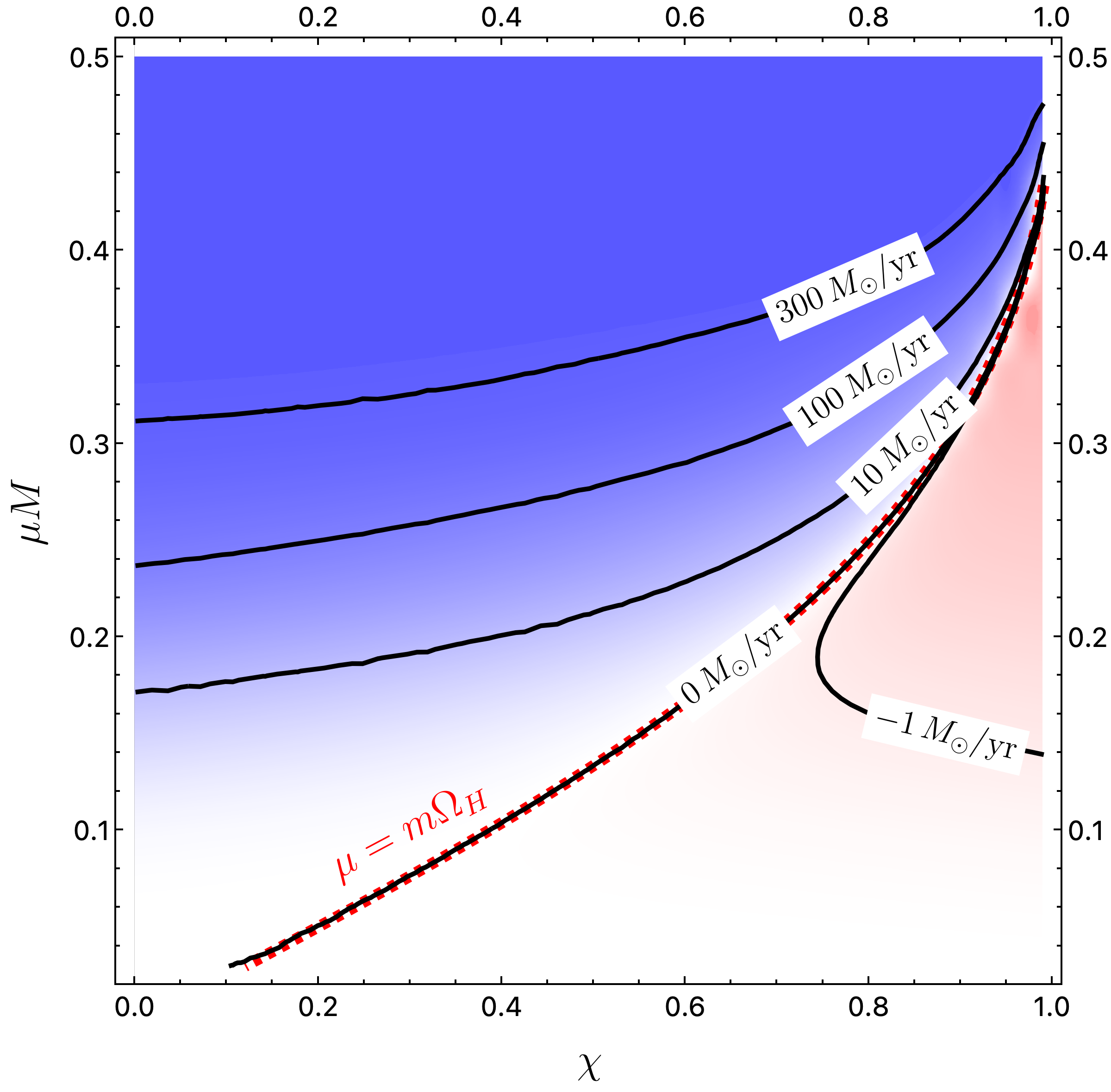}
    \caption{Contours of the \bh{} mass accretion rate $\dot{M}$ in the $(\chi, \mu M)$ plane for the dominant superradiant mode, $(s, \ell, m) = (-1, 1,1)$, calculated for a $M=10^9\msun$ \bh{} in a dark matter environment with $\rho_c = 10 \msun/\pc^3$. Deep in the particle ($\mu M \gtrsim 1$) and wave ($\mu M < 1$) regimes, the effects of \bh{} rotation are suppressed, but a superradiant regime with $\dot{M} < 0$ is present, with the rate surpassing $\dot{M} \sim 1 \msun/\yr$ for $\mu M \sim 0.3$, $\chi> 0.8$. The superradiance cutoff $\mu = m \Omega_H$ is shown as red, dashed line.}
    \label{fig:Mdot_contours_s-1l1m1}
\end{figure}

In Figure~\ref{fig:Mdot_contours_s-1l1m1}, we plot contours of constant mass accretion rate in the spin-mass $(\chi, \mu M )$ plane for the $(s,\ell, m) = (-1,1,1)$ mode.
The superradiance threshold, $\mu = m \Omega_H$, is indicated as a red dashed line. 
We use blue to indicate the region where $\dot{M}>0$ and red to indicate $\dot{M} <0$, with the opacity indexed to the magnitude of the accretion rate. We see that the superradiance threshold coincides with a hard break in the phenomenology, exactly as expected;
for $\mu > m \Omega_H$, the \bh{} grows in mass ($\dot{M} > 0$), while for $\mu < m \Omega_H$, the \bh{} mass ``shrinks" ($\dot{M} < 0$). 
To understand how the field can
decrease the \bh{} mass,
we recall that the Kerr \bh{'s} total (Komar or ADM) mass can be expressed as
\begin{equation}
    M^2 = M^2_\text{irr} + \frac{1}{4} \frac{J^2}{M^2_\text{irr}},
\end{equation}
where $M_\text{irr}$ is the \bh{'s} irreducible mass and $J$ is the angular momentum~\cite{Wald:1984rg}. In a time $\delta t$, the field extracts $\delta J$ in angular momentum from the \bh{} while leaving $M_\text{irr}$ unchanged at linear order.
It thus changes the \bh{'s} total mass by an amount $\delta M = \Omega_H \delta J$. In the superradiant case, the field extracts angular momentum from the \bh{}, so $\delta J<0$, and thus $\delta M < 0$. While not plotted, the qualitative features of the equivalent plots for the $(0,1,1)$ and $(1,1,1)$ modes are the same. The main difference is that the superradiant-regime mass extraction rates are much smaller for these modes than for $(-1,1,1)$. We explore this in more detail below. 

\begin{figure}[!ht]
    \centering
    \includegraphics[width=0.9\linewidth]{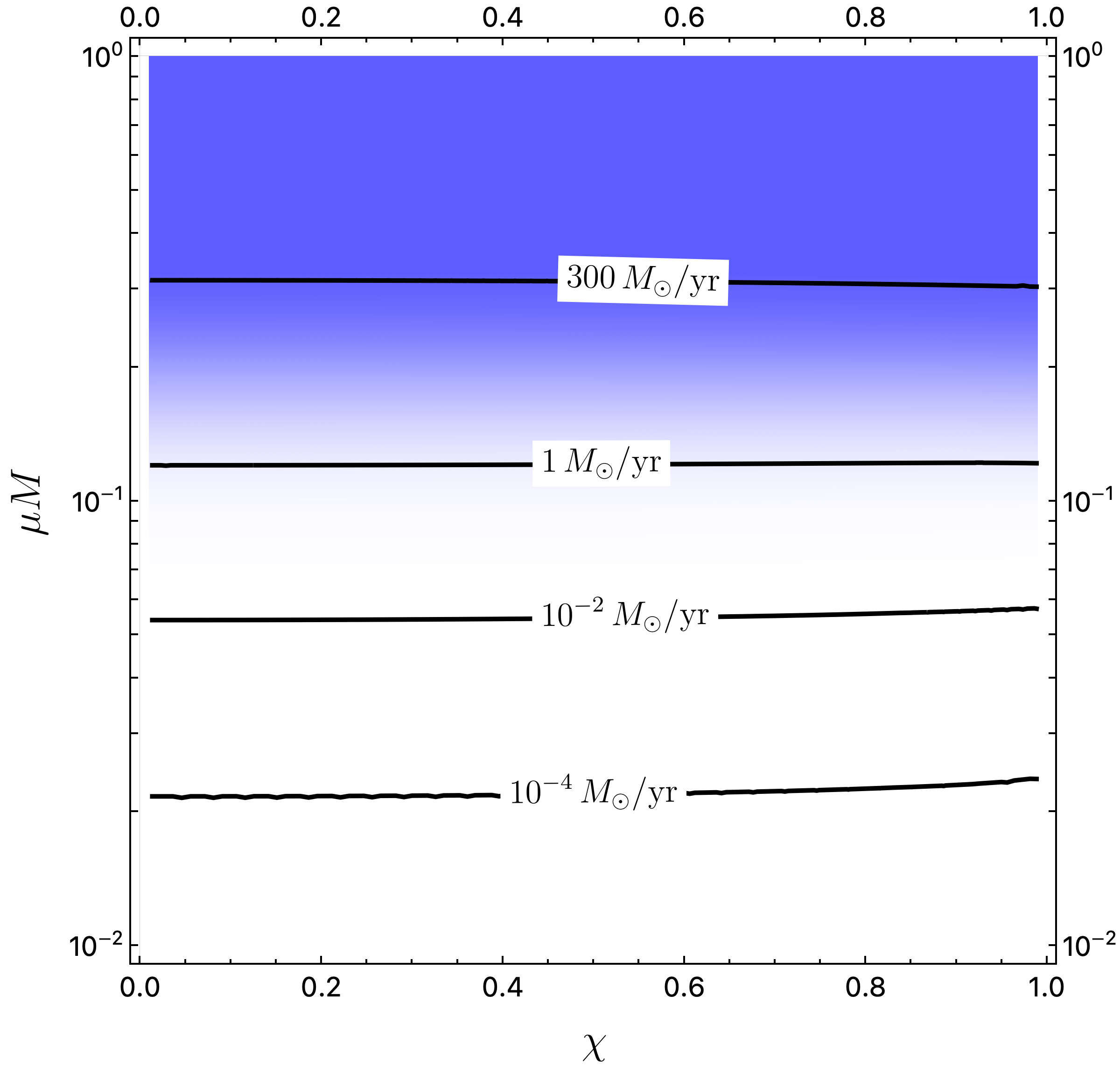}
    \caption{Contours of the \bh{} mass accretion rate $\dot{M}$ in the $(\chi, \mu M)$ plane for the monopole mode, $(s, \ell, m) = (1,0,0)$, calculated for a \bh{} of mass $M=10^9\msun$ in a dark matter environment with $\rho_c = 10 \msun/\pc^3$.
    There is no superradiant regime since $m = 0$.
    Rotation has very little effect on the accretion rate.}
    \label{fig:Mdot_contours_s1l0m0}
\end{figure}

In Figure~\ref{fig:Mdot_contours_s1l0m0} we show an equivalent plot of the mass accretion rate contours in the spin-mass plane for the monopole $(1,0,0)$ mode.
Since $m=0$, there is no superradiant regime and the superradiance threshold is not plotted.
We also see that the mass accretion rates of the monopole mode remain nearly constant as the \bh{} spin $\chi$ increases.
Thus, the Schwarzschild monopole accretion rates are essentially valid for any \bh{} spin. 

\begin{figure}
    \centering
    \includegraphics[width=0.95\linewidth]{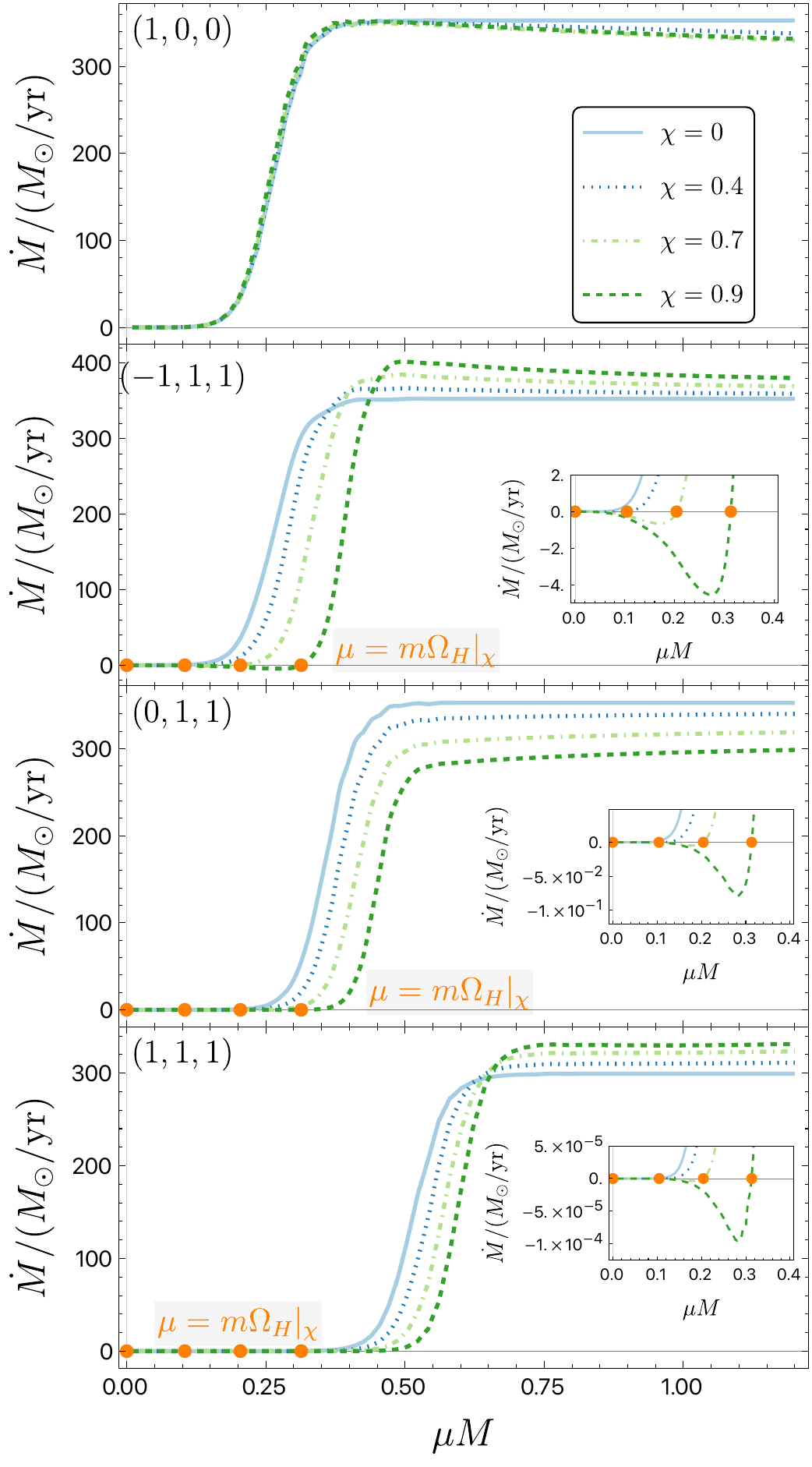}
    \caption{\bh{} mass accretion rate $\dot{M}$, for a \bh{} of mass $M=10^9\, \msun$ in a $\rho_c = 10\, \msun/\pc^3$ dark matter environment, plotted against $\mu M$ for several values of the \bh{} spin $\chi$. 
    From top to bottom, the panels show the modes $(s,\ell,m)=(1,0,0)$, $(-1,1,1)$, $(0,1,1)$, and $(1,1,1)$. For the $\ell=m=1$ modes, the curves all cross zero at the point $\mu = m \Omega_H$, giving rise to a superradiant regime for lower $\mu$ values, in which the \bh{} loses mass. Each of these plots includes an inset, showing the curves' superradiant-regime behavior in more detail. For $\mu > m \Omega_H$, the mass accretion rate levels off to between $300\, \msun/\yr$ and $400\, \msun/\yr$ for all modes.}
    \label{fig:Mdot_line_plots}
\end{figure}

Quantitative features of these plots are easier to extract if we reduce their dimension.
In Figure~\ref{fig:Mdot_line_plots}, we plot the mass accretion rate as a function of just the mass parameter, plotted for several fixed \bh{} spins in the range $\chi \in \{0,0.4,0.7,0.9\}$.
From top to bottom, the panels show the modes 
$(s,\ell,m)=(1,0,0)$, $(-1,1,1)$, $(0,1,1)$, and $(1,1,1)$. 
For modes with a superradiant regime, we indicate the superradiant threshold 
with orange points,
and show a magnification of this region in the inset. 

Here, we see that all modes asymptote to an particle-regime accretion rate of approximately
\begin{equation}
    \label{eq:mdot_estimate_particle}
    \dot{M} \approx \left(400 \frac{\msun}{\yr} \right) \left(\frac{M}{10^9 \msun}\right)^2 \left(\frac{\rho_c}{10 \msun/\pc^3}\right)\, ,
\end{equation}
closely coinciding with the Schwarzschild particle-regime accretion rates we found in Paper~I. We may define the mass accretion timescale as
\begin{equation}
    \label{eq:mass_acc_timescale_def}
    \tau_\text{acc} \coloneq \frac{M}{|\dot{M}|}\, ,
\end{equation}
which corresponds to the timescale of the \bh{'s} mass change via interaction with its vector dark matter environment.
Using Eqs.~\eqref{eq:mdot_estimate_particle} and~\eqref{eq:mass_acc_timescale_def}, we see that in the particle regime, this timescale is given roughly by
\begin{equation}
    \label{eq:mass_acc_timescale_estimate}
    \tau_\text{acc} \sim \left(10^7\, \yr\right)  \left(\frac{M}{10^9 \msun}\right)^{-1} \left(\frac{\rho_c}{10 \msun/\pc^3}\right)^{-1}\, .
\end{equation}
This is longer than a Hubble time for \bh{s} of less than $10^6\, \msun$, 
but for supermassive \bh{s} with $M \gtrsim 10^8\, \msun$, this timescale is comparable to or shorter than galactic evolution timescales. 

In the top panel of Figure~\ref{fig:Mdot_line_plots}, we see that the $(1,0,0)$ curves match almost exactly for all spins, except at high $\mu M$, 
where the accretion rate $\dot{M}$ is slightly suppressed at high spin.
The lower three panels of Figure~\ref{fig:Mdot_line_plots} show 
the mass acretion rate
for the three fundamental $m=1$ modes. 
For nonzero spins, superradiance introduces a region with negative accretion rate, $\dot{M} < 0$, in the wave regime,
which indicates the extraction of \bh{} mass when
$\mu < m \Omega_H$; 
see the insets in Figure~\ref{fig:Mdot_line_plots}.
We see that the $(-1,1,1)$ rates dip the deepest below zero, as expected,
indicating the largest extraction of \bh{} mass.
We also find that the extraction rates due to the $(-1,1,1)$ mode is about two orders of magnitude larger than the $(0,1,1)$ mode and four orders of magnitude larger than the $(1,1,1)$ mode.
In contrast, in the particle regime, the \bh{} spin has a much smaller effect on the behavior of these modes.
It only slightly enhances the mass accretion rate for the  $(-1, 1,1)$ and $(1,1,1)$ modes while slightly suppressing it for $(0,1,1)$. 

\bh{} mass loss via superradiance is maximized for high spin
$\chi$
and for $\mu M$ values that are just below the 
superradiant
threshold. 
Considering parameters near the maximum \bh{} mass extraction,
about $\mu M \approx 0.3$ and $\chi > 0.9$ for the $(-1,1,1)$ mode
(see Figs.~\ref{fig:Mdot_contours_s-1l1m1}
and~\ref{fig:Mdot_line_plots}),
the \bh{} loses mass into the cloud 
at a rate
\begin{equation}
    \label{eq:mdot_estimate_superrad}
    \dot{M} \approx -\left(10 \frac{\msun}{\yr} \right) \left(\frac{M}{10^9 \msun}\right)^2 \left(\frac{\rho_c}{10 \msun/\pc^3}\right)\, .
\end{equation}

While small relative to the $\sim400 \msun/\yr$ accretion rates for
a black hole of the same mass and spin
found in the particle regime,
and as we will see, small relative to the growth rate of the superradiant instability, this is large relative to its scalar-field counterpart, as we show in the next section.

\subsubsection{Comparison with scalar results}

To draw out the implications of our results, we can compare these accretion rates with those of a massive scalar field in an analogous configuration. For this, we may use the mass accretion formula from Ref.~\cite{Hui:2022sri}.

We consider a scalar field $\Phi$ with mass
parameter
$\mu$ around a Kerr \bh{}, populated in a single mode $(\ell, m)$ with a definite complex frequency $\omega \in \mathbb{C}$. The field may be expressed as
\begin{equation}
    \label{eq:scalar_ansatz}
    \Phi = R(r) S(\theta) e^{-i \omega t} e^{i m \phi}\, ,
\end{equation}
which separates the Klein-Gordon equation into a system of \ode{s}. The field's energy-momentum tensor has a simpler structure than its Proca counterpart, which allows us to greatly simplify the mass accretion rate formula in Eq.~\eqref{eq:mass_accretion_rate_general_formula} by taking the constant-$r$ integration surface to be just outside the horizon, $r \to r_+$~\cite{Hui:2019aqm, Hui:2022sri}. This yields
\begin{equation}
    \label{eq:scalar_Mdot}
    \dot{M}_\text{Scal} = 4 M r_+ \left( |\omega|^2 - m \Omega_H \omega_R \right) |R_+|^2\, ,
\end{equation}
where $|R_+|^2 \coloneq \lim_{r \to r_+} |R(r)|^2$. 
In the regime $r \gg r_+$, the 
scalar
field's density is
\begin{equation}
    \label{eq:scalar_density_gen_expr}
    \rho_\text{Scal} \approx \left( |\omega|^2 + \mu^2 \right) |\Phi|^2 + |\partial_i \Phi|^2\, .
\end{equation}
In the non-relativistic regime, where we have $\omega \approx \mu$ and the spatial gradient vanishes, the scalar's density reduces to $\rho_{\rm{Scal}} \approx 2 \mu^2 |\Phi|^2$.
We wish to normalize the field to its averaged density at
the edge of the black hole's sphere of influence,
$r = r_c \sim 10^6 M$.
For our modes of interest, we have
$\rho_c = \rho_{\rm{Scal}}(r_{c})$ with
\begin{align}
\label{eq:scalar_rhoc_expression}
\rho_c
    \approx \mathcal{N} \mu^2 |R_c|^2
\,,\quad
\mathcal{N}\coloneq\begin{cases}
        \frac{1}{2 \pi}  & \ell = m = 0\\
        \frac{3}{8 \pi} & \ell = m =1
    \end{cases}\, ,
\end{align}
where $|R_c|^2 \coloneq |R(r_c)|^2$. 
To normalize the mass accretion rate, we take $|R_+|^2 = \frac{|R_+|^2}{|R_c|^2}|R_c|^2$ in Eq.~\eqref{eq:scalar_Mdot}, then express the latter $|R_c|^2$ in terms of $\rho_c$ using Eqs.~\eqref{eq:scalar_ansatz} and~\eqref{eq:scalar_density_gen_expr}.
This yields the $\omega = \mu$ expression
\begin{equation}
    \label{eq:scalar_Mdot_nonrel}
    \dot{M}_\text{Scal} = \mathcal{N} \frac{M r_+ \rho_c}{\mu} \left(\mu - m \Omega_H\right) \frac{|R_+|^2}{|R_c|^2}\, .
\end{equation}
We may pull
results for the fraction
$|R_+|^2/|R_c|^2$ 
from Eq.~(4.1) of Ref.~\cite{Hui:2022sri}.

\begin{figure}[!ht]
    \centering
    \includegraphics[width=0.95\linewidth]{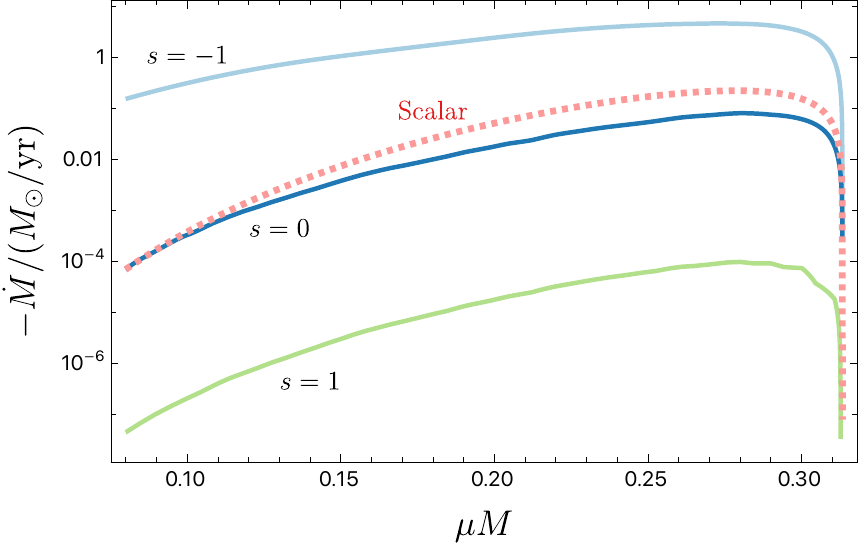}
    \caption{\bh{} mass extraction rate, $-\dot{M}$, for a \bh{} of mass $M=10^9 M_\odot$ and spin $\chi = 0.9$ accreting from an $\ell=m=1$
    dark matter cloud with density $\rho_c = 10\msun/\pc^3$ at $r=r_c$.
    The solid lines show the 
    extraction rates 
    due to a Proca-field cloud
    for each polarization $s = (-1,0,1)$, while the dotted line shows the $\ell=m=1$ scalar-field extraction rate. 
    We see that the $s=0$ polarization
    of the vector field, with total angular momentum $j=\ell+s=1$, matches the 
    $j=\ell=1$
    scalar field up to an $\mathcal{O}(1)$ factor, while the $s=-1$, $j=0$ mode is enhanced and the $s=1$, $j=2$ mode is suppressed.}
    \label{fig:Mdot_superrad_scalar_compare}
\end{figure}

To compare results
of the scalar-field and Proca-field accretion rates
in the superradiant regime, we select the ``intermediate'' scalar-field solution for $|R_+|^2/|R_c|^2$
given in Eq.~(4.1) of Ref.~\cite{Hui:2022sri}.
This allows us to evaluate
the \bh{'s} mass accretion/extraction rate due to a scalar field, 
given in Eq.~\eqref{eq:scalar_Mdot_nonrel},
for a range of values in the crucial regime $\mu M \lesssim 0.4$.
In Figure~\ref{fig:Mdot_superrad_scalar_compare} we plot the mass extraction rate for a \bh{} of spin $\chi=0.9$.
The scalar results are superimposed onto those for each vector polarization's dominant superradiant mode.
We see that the scalar closely follows the $s=0$ vector-field mode, while the $s=1$ and $s=-1$ modes are parametrically suppressed and enhanced, respectively, relative to the scalar. 
This is precisely the behavior we expected from angular momentum addition arguments, and these ratios match those between the superradiant instability growth rates for the vector-field polarizations~\cite{Dolan:2018dqv}. 

\subsection{Angular momentum accretion}

In this work, we have considered Proca-field perturbations of the Kerr spacetime which are populated only in individual modes 
$(s,\ell, m)$ with definite $\omega$. 
In more general terms, a field populated in a mode of definite $(\omega, m)$, together with the isometry generators $(\partial_t, \partial_\phi)$ (Killing vectors), 
forms a representation of the Kerr spacetime isometry group.
As originally argued in Ref.~\cite{Bekenstein:1973mi}, since the Proca field's total mass $M_A$ and angular momentum $J_A$ are each built from the
same fixed number of the quanta $\omega$ and $m$, respectively, 
any change in the former, $\delta M_A$,  corresponds to a 
change in the latter, $\delta J_A$, multiplied by the ratio $m / \omega$, so $\delta J_A = m/\omega\,\delta M_A$. 
The \bh{} then absorbs these quanta, and its mass and angular momentum change accordingly. 
Upon a more careful analysis, given in Appendix~\ref{sec:Mdot_to_Jdot}, we find that the precise relation implied by this mechanism is
\begin{equation}
    \label{eq:bh_Jdot_to_Mdot}
    \dot{J} = \frac{m}{\omega_R} \dot{M}\, ,
\end{equation}
where $J$ is the \bh{'s} total angular momentum. This is a well accepted result in the literature, but since it does not follow obviously from the Proca field's energy-momentum tensor, we derive it explicitly in Appendix \ref{sec:Mdot_to_Jdot}. 

\begin{figure}
    \centering
    \includegraphics[width=0.95\linewidth]{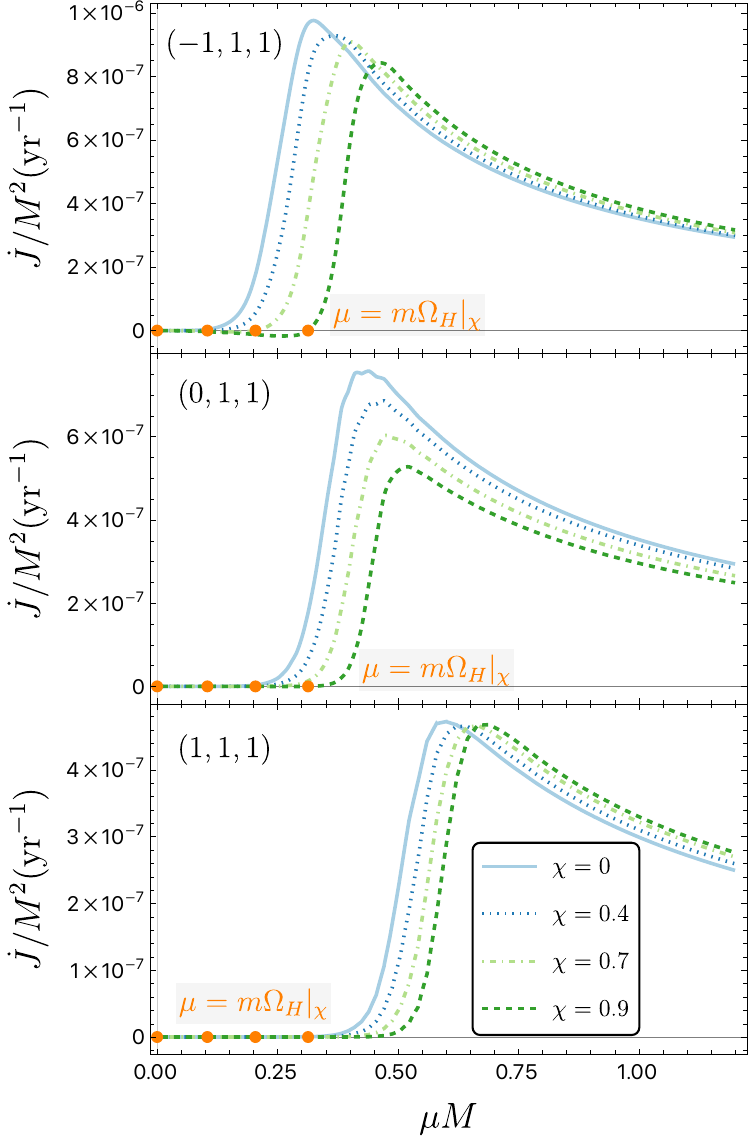}
    \caption{\bh{} angular momentum accretion rate $\dot{J}$, for a \bh{} of mass $M=10^9\, \msun$ in a dark matter environment with $\rho_c = 10\, \msun/\pc^3$, plotted against $\mu M$ for several values of the spin $\chi$. 
    From top to bottom, the panels show the modes $(s,\ell,m)=(-1,1,1)$, $(0,1,1)$, and $(1,1,1)$. The units are $M^2/\yr$, so the inverses of
    the y-axis values roughly give the timescale on which the dimensionless spin $\chi$ changes. 
    The curves all cross zero at the point $\mu = m \Omega_H$, giving rise to a superradiant regime for 
    $\mu< m\Omega_{H}$,
    in which the \bh{} loses angular momentum. 
    For $\mu > m \Omega_H$, the \bh{} gains angular momentum with time, achieving its peak rate at a point between
    $\mu M=0.2$ and $\mu M = 0.7$.}
    \label{fig:Jdot_line_plots}
\end{figure}

From Eq. \eqref{eq:bh_Jdot_to_Mdot}, we can calculate angular momentum accretion rates using our
results for the mass accretion rates.
In Figure~\ref{fig:Jdot_line_plots} we plot the angular momentum accretion rates against the mass parameter for the three $\ell=m=1$ modes at \bh{} spins $\chi \in \{0,0.4,0.7,0.9\}$.
The superradiance threshold for each curve is again shown as an orange point.
Note that $\dot{J}$ vanishes for the monopole mode, since $m=0$.
At low $\mu M$, the same behavior as for $\dot{M}$ persists; 
$\dot{J}$ is negative in the superradiant regime, $\mu < m \Omega_H$, with the $(-1, 1, 1)$ mode reaching the 
largest negative values.
In the accretion regime, $\mu > m \Omega_H$, 
$\dot{J}$ rises steeply to a single maximum, beyond which it falls off with the $\dot{J} \sim (\mu M)^{-1}$ tail inherited from Eq. \eqref{eq:bh_Jdot_to_Mdot}. 
The precise location of the peak depends on both the spin projection $s$ and the \bh{} spin $\chi$, with an increase in either quantity shifting the peak to a higher $\mu M$ value. For all modes and \bh{} spins considered here, the peaks lie between $\mu M=0.2$ and $\mu M = 0.7$. 

The \bh{} angular momentum accretion 
rates are most easily understood in association with the dimensionless spin parameter $\chi = J/M^2 \in [0,1)$.
Its time derivative is
\begin{align}
    \label{eq:chidot_expression}
    \dot{\chi} = \frac{\dot{J}}{M^2}-2\chi \frac{\dot{M}}{M}
\, .
\end{align}
The curves in Figure~\ref{fig:Jdot_line_plots} give the first term in this expression for a $M = 10^9\, \msun$ \bh{}, while the second can be derived from Figure~\ref{fig:Mdot_line_plots}. 
The $(-1,1,1)$ mode, for instance, at its accretion-regime peak, somewhere between $\mu M = 0.2$ and $\mu M = 0.5$ for all spins $\chi$, reaches $\dot{J}/M^2 \sim~10^{-6}\, \yr^{-1}$. The second term in Eq.~\eqref{eq:chidot_expression}, however, is roughly of magnitude $1/\tau_\text{acc} \sim 10^{-7}\, \yr^{-1}$.
A similar discrepancy persists for the other modes. 
Thus, near the peaks in Figure~\ref{fig:Jdot_line_plots}, 
the first term in Eq.~\eqref{eq:chidot_expression} and we may consider only $\dot{\chi} \approx \dot{J}/M^2$.

Using these results, we may define a timescale associated with \bh{} spin change, and estimate it for the entire particle regime as
\begin{align}
    \label{eq:spin_timescale}
    \tau_\text{spin,acc} ={}& \frac{\chi}{|\dot{\chi}|}\\\nonumber
    \sim{}& \left(10^6\, \yr\right) (\mu M)\left(\frac{M}{10^9\msun}\right)^{-1} \left(\frac{\rho_c}{10 \msun/\pc^3}\right)^{-1}\, .
\end{align}
Like with mass accretion, we find that the larger 
the \bh{} mass
$M$ is, the more the field affects the \bh{} spin. 
For a  supermassive \bh{} 
of mass $M=10^9 \msun$
in a $\rho_c=10 \msun/\pc^3$ environment with 
$\mu > \Omega_H$, if the system were anywhere near $\mu M \sim 0.4$, then the \bh{} would be spun up to near the superradiance threshold in $10^6$ years, which is very short compared to galactic evolution timescales.
This timescale is longer by a factor $(10^9\, \msun)/M$ for smaller \bh{s}, exceeding a Hubble time for \bh{s} smaller than $M \sim 10^5\, \msun$, making angular-momentum accretion most relevant for the largest \bh{s}, similarly to mass accretion.
Thus, a dense dark matter cloud with a significant amount of angular momentum might impact a \bh{'s} spin more than its mass. 

Like for mass accretion, the angular momentum accretion rate is greatly suppressed in the wave regime. However, in its superradiant region, the field can still spin down the \bh{} at an appreciable rate.
In the $M=10^9\, \msun$ and $\rho_c = 10\, \msun/\pc^3$ scenario,
the $s=-1$ superradiant-regime spin-down timescale reaches as low as 
$\tau_\text{spin,acc} \sim 10^8\, \yr$, though it is far longer for \bh{} spins close to the superradiance threshold, much like the superradiant instability timescale $\tau_\text{spin,ins}$~\cite{Pani:2012bp, Pani:2012vp, Dolan:2018dqv}.

In all regimes, the ratio of the spin to mass accretion timescales for all $m>0$ modes is 
$\tau_\text{spin,acc} / \tau_\text{acc} \approx \mu M/m$, found by eliminating $\dot{M}$ and $\dot{J}$ from Eq.~\eqref{eq:bh_Jdot_to_Mdot} using the definitions of the two timescales, given in Eq.~\eqref{eq:mass_acc_timescale_def} and Eq.~\eqref{eq:spin_timescale}. 
Thus, for $\mu M \gg 1$, the field spins up the \bh{} by a negligible amount compared to its mass growth contribution
in the same time interval.
Angular momentum accretion is relevant 
only for $\mu M \sim 1$.


\subsection{Comparison with superradiant instability growth rate}



In this work, we consider solutions for which the field's frequency $\omega$ is held constant to the real value $\omega=\mu$, modeling a \bh{} in equilibrium with a cold dark matter bath. 
The field's 
amplitude thus does not change with time, whether in the superradiant or in the accretion regime.
A generic perturbation will excite some superposition of the system's quasi-normal modes and quasi-bound states---modes with time-dependent amplitudes---but most of these have $\omega_I < 0$ and will decay,
thus returning the system to our equilibrium configuration. 

With \bh{} rotation, however, any massive bosonic field is susceptible to unstable growth due to the superradiant instability; 
any quasi bound state with $\omega_R < m \Omega_H$ will have $\omega_I > 0$. 
Rather than settling to an equilibrium configuration, then, the field would blow up, with its density schematically going as
$\rho \sim |A|^2+|\partial A|^2\propto e^{2 \omega_I t}$.  
Thus, unless the \bh{} equilibrates with its environment on a timescale faster than $\sim1/\omega_I$, then it would be unable to form the equilibrium profile considered in this work before being pushed to the superradiance threshold by the superradiant instability.
Thus, the question at hand is whether the environment equilibration timescale exceeds the superradiant instability timescale.
We can answer this using both results from this work and results from the literature.

To compare the timescales, we take as an example an initially highly spinning, supermassive \bh{} with  $M=10^9 \msun$ and $\chi = 0.9$,
surounded by a Proca field with mass parameter $\mu M \approx 0.3$.
Then, the fundamental co-rotating $s=-1$ mode has a superradiant instability timescale of $\tau_\text{ins} = 1/\omega_I \sim 1\, \yr$~\cite{Dolan:2018dqv}. 
Conversely, taking the accretion timescale $\tau_\text{acc} = M/|\dot{M}_\text{acc}|$, defined in Eq.~\eqref{eq:mass_acc_timescale_def}, as a proxy for the equilibration timescale, 
and pulling $\dot{M}_\text{acc}$ from Eq.~\eqref{eq:mdot_estimate_superrad}, 
the \bh{} equilibrates with its dark matter environment on a timescale of $\tau_\text{acc} \sim 10^8\, \yr$. 


Clearly, the superradiant instability is much faster, so the \bh{} will be spun down to near the superradiance threshold long before it equilibrates with its dark matter environment. 
This is true for most 
parameters
$(\chi,\mu)$ in the superradiant regime, except for those very close to its threshold $\mu = m \Omega_H$. 
Additionally, relative to the \bh{} mass, the timescales scale as $\tau_\text{ins} \propto M$~\cite{Pani:2012bp} and $\tau_\text{acc} \propto 1/M$, as seen in Eq.~\eqref{eq:mass_acc_timescale_estimate}. Thus, for \bh{s} of mass $M < 10^9 \msun$, this discrepancy is even larger. 
A similar story persists when considering \bh{} spin; the shortest possible \bh{} spin-down timescale from vector dark matter interaction, $\tau_\text{spin,acc}$, defined in Eq.~\eqref{eq:spin_timescale}, is several orders of magnitude larger than the superradiant instability spin-down timescale, $\tau_\text{spin,ins}$, for virtually any astrophysical \bh{} in the superradiant regime.
The superradiant instability therefore precludes any astrophysical \bh{} from establishing an equilibrium with its vector dark matter environment while in the superradiant regime.
Although such equilibrium configurations are theoretically possible as shown in this work, they should be considered unstable to small perturbations. 
The regime in which equilibrium energy-momentum exchange between a \bh{} and its vector dark matter environment is astrophysically relevant is rather the non-superradiant regime,
$\mu > m \Omega_H$, discussed in previous sections. 


\section{Summary and discussion}
\label{sec:discussion}

In this work, we showed that a massive vector perturbation of a Kerr \bh{} forms a density spike when allowed to accrete endlessly from 
a bath of fixed asymptotic density---representing a rotating \bh{} in a universe filled with a vector dark matter condensate. 
The spike is morphologically similar to the Schwarzschild spike we found in Paper~I for all \bh{} spins; 
outside the horizon's immediate
environment
it has an averaged density profile $\rho \sim r^{-3/2}$, punctuated by regular standing-wave nodes for the wave regime ($\mu M < 1$). 
Near the horizon of a spinning \bh{}, the monopole $(1,0,0)$ mode's density profile is little affected by the \bh{} spin.
However, corotating 
modes
experience some distortion due to frame dragging, namely, developing a sharper central spike and nearby regions of reduced density. 
The principal deviation from the Schwarzschild behavior we found was in the rates of \bh{} mass and angular momentum change due to accretion from the field. 
For mass accretion in the particle regime ($\mu M \gtrsim 1$)
we found behavior essentially the same as the Schwarzschild results of Paper I for all \bh{} spins. 
In the wave regime, however, a new superradiant
region emerged for $\mu < m \Omega_H$, in which both
the mass and angular momentum accretion rates,
$\dot{M}$ and $\dot{J}$,
are negative. 
In other words, in the superradiant regime, the dark matter spike in its steady-state configuration steadily extracts mass and angular momentum from the \bh{}, rather than accreting onto it. However, the timescales of both mass and angular momentum extraction by this process
are far longer than that of the superradiant instability to which the system is also susceptible for $\mu < m \Omega_H$ if its quasi-bound states are perturbed.
Thus, if the quasi-bound states were present, as for any physical \bh{}, then the superradiant instability 
would be the dominant process.

We also presented several novel mathematical results in this work.
We extended a result by Dolan~\cite{Dolan:2018dqv} that the \lfkks{} method recovers the \vsh{} $s=0$ radial equation in the limit 
$\chi \to 0$, explicitly showing that the $s=0$ \vsh{} expression appears as the leading-order term in the small-$a\mu$ expansion of the $s=0$ \lfkks{} expression for $A_\mu$ (see Eq. \eqref{eq:fkks_to_vsh_s=0}). Pursuant to our accretion-rate study, we derived generic expressions for the accretion rates $\dot{M}$ and $\dot{J}$ in Appendix \ref{sec:acc_rates_defs}. We then explicitly derived the standard relation between them, $\dot{J}/\dot{M} = m/\omega_R$, showing that the ratio only holds exactly in the case $\omega \in \mathbb{R}$ for the Proca field, with an additional term appearing for $\omega_I \neq 0$. This latter result has not been shown elsewhere in the literature, to our knowledge.

These results open the door to a quantitative study of how 
vector
dark matter accretion mixes with the superradiant instability, affecting the end-state of the vector-field cloud produced by the latter.
This has been done in linear perturbation theory for the scalar case, where it was found that dark matter accretion can feed a superradiant mode during its growth, leading to a more massive superradiant cloud---a phenomenon dubbed ``over-superradiance''~\cite{Hui:2022sri}.

Here, we showed that in the superradiant regime, the $(s,\ell,m) = (-1,1,1)$ vector-field mode causes the \bh{} to lose mass at a rate 2-3 orders of magnitude faster than its scalar-field counterpart, mirroring its superradiant instability growth rate, which is larger than the scalar's by a similar factor~\cite{Pani:2012bp, Pani:2012vp,Witek:2012tr,Dolan:2018dqv}.
Thus, the process of over-superradiance may be significantly altered in the vector case, changing the final cloud mass.
We will leave a thorough investigation of this, however, to future work.

While this work illuminates the contribution of \bh{} rotation to the steady-state behavior of this system, 
it does not
capture the dynamical behavior of the field 
when perturbed from this configuration or the effect of the dark matter spike on the spacetime metric.
To understand these, we must turn to a full numerical-relativity simulation of the system. 
For scalar fields, the accretion has been studied for 
single nonrotating~\cite{Clough:2019jpm} or rotating \bh{s}~\cite{Bamber:2020bpu}, 
and binary \bh{s}~\cite{Bamber:2022pbs, Aurrekoetxea:2023jwk, Aurrekoetxea:2024cqd, Cheng:2025wac}.
While Proca fields have frequently been simulated in \bh{} spacetimes, especially in the context of the superradiant instability~\cite{Witek:2012tr,East:2017mrj, East:2017ovw, East:2018glu, Xin:2024trp}, a direct numerical-relativity counterpart of the vector dark matter accretion model from this work has not been studied. 
This could be done along much the same lines as the scalar-field studies,
though more attention would need to be paid to initial data, both due to the Proca field's extra degrees of freedom and due to the more complicated structure of its constraints.
We plan to adapt the formalism used for Proca-field simulations in Ref.~\cite{Zilhao:2015tya}.
After solving these technical problems, however, one could address questions like the timescale on which the spike forms, which polarizations become most prominent, and whether small perturbations disrupt the spike near or beyond the superradiance threshold.

\section*{Acknowledgements}

We thank 
S.~Dolan and
H.~O.~Silva
for insightful discussions and comments.
H.~W. acknowledges support provided by the National Science Foundation under NSF Award 
No.~OAC-2411068 and No.~PHY-2409726.
F.~H. acknowledges support from the National Science Foundation Graduate Research Fellowship Program under Grant No. DGE 21-46756. 

We acknowledge the Centro de Ciencias de Benasque Pedro Pascual, Spain, for the hospitality during the ``New Frontiers in Strong Gravity IV'' workshop, during which this work was completed.

\appendix

\section{Near-horizon asymptotics of radial equation}
\label{sec:radial_eq_asymptotics}

Here, we explicitly work out the asymptotics of the radial equation \eqref{eq:fkks_radial_eq_final}. 
In the near-horizon regime, we have several useful limits, recalling the definitions in Eqs.~\eqref{eq:Delta_def} and~\eqref{eq:aux_vars_defs}, 
\begin{subequations}
\begin{align}
    \lim_{r \to r_+} \Delta ={}& 0\\
    \lim_{r \to r_+} K_r^2 ={}& K_{r_+}^2 = (2Mr_+)^2(\omega -m \Omega_H)^2\, ,
    \label{eq:Kr_rp_limit}
\end{align}
\end{subequations}
where $\Omega_H = \frac{a}{2 M r_+}$
is the angular velocity of the horizon. 
Several terms in Eq.~\eqref{eq:fkks_radial_eq_final} can be immediately eliminated in the limit $r \to r_+$, and the equation reduces to
\begin{equation}
\label{appeq:RadODENearHor}
\left[q_r \Delta\frac{d}{dr}\left(\frac{\Delta}{q_r} R'(r)\right)\right] + K_{r_+}^2 R(r)
= 0
\,.
\end{equation}
Applying the limit to the derivative term requires more care. To make sense of it, we switch to the tortoise coordinate, $r \to r_*$, implicitly defined by $\dif r_* = \frac{r^2 + a^2}{\Delta} \dif r$. 
In the resulting lengthy expression, the $R'(r_*)$ coefficient vanishes,
while the $R''(r_*)$ term's coefficient 
becomes
$(2M r_+)^2$.
Dividing out the latter factor and taking the limit in Eq. \eqref{eq:Kr_rp_limit}, 
Eq.~\eqref{appeq:RadODENearHor}
reduces to
\begin{equation}
    R''(r_*) + (\omega - m \Omega_H)^2 R(r_*) = 0\, .
\end{equation}
This has the general solution
\begin{equation}
    R(r_*) = C_1 e^{i \left(\omega - m \Omega_H\right) r_*} + C_2 e^{-i \left(\omega - m \Omega_H\right) r_*}\, ,
\end{equation}
which are ingoing and outgoing spherical waves. 
To impose regularity on the horizon, we must discard all outgoing modes on the horizon, meaning we seek solutions such that $C_1 = 0$ near the horizon. From the imaginary part of the exponent, we notice, however, that the ``ingoing" solution becomes outgoing in the regime
\begin{equation}
    \omega_R < m \Omega_H
\,,
\end{equation}
This is precisely the classic superradiance condition, and the calculation in this section can be seen as a derivation of it. We should expect solutions in this regime to exhibit signs of superradiant scattering.

\section{Mass and angular momentum accretion rates}

\label{sec:acc_rates_defs}

\begin{figure}
    \centering
    \includegraphics[width=0.9\linewidth]{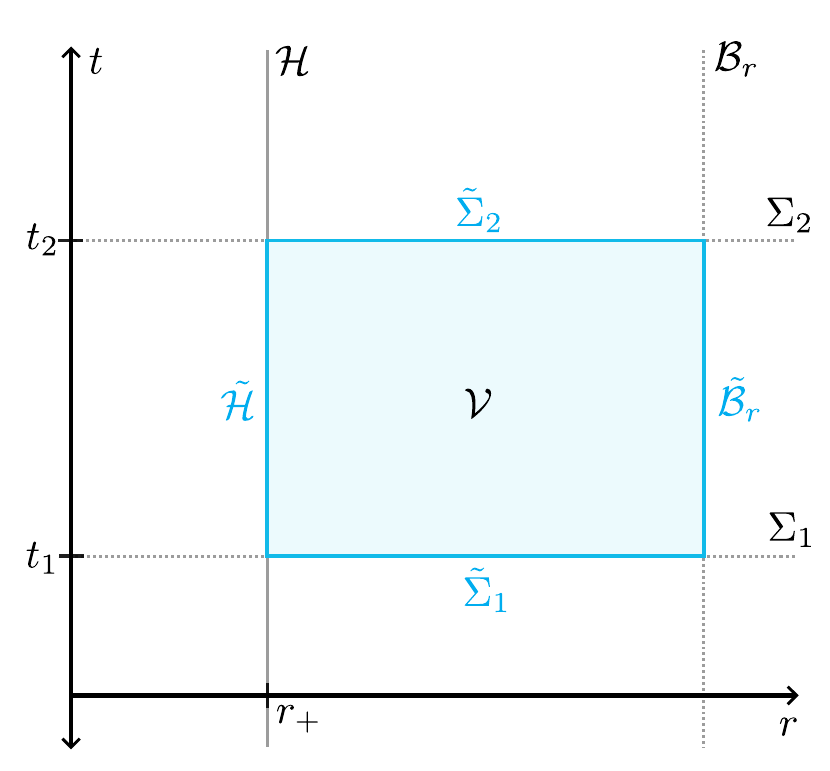}
    \caption{The surfaces $\mathcal{H}$, $\mathcal{B}_r$, and $\Sigma_{(i)}$, and their truncations (notated by tildes) to the boundary of the region $\mathcal{V}$, shown in the $(t,r)$ plane. The outer horizon lies at $r=r_+$, and the two spatial slices lie at $t=t_1$ and $t=t_2$, as indicated.
    }
    \label{fig:accretion_surfaces_sketch}
\end{figure}

In this section, we derive the expressions used for the mass and angular momentum accretion rates in the text.
The Kerr metric manifestly has the Killing vectors
\begin{align}
    t^\mu \partial_\mu = \partial_t\,, \quad \phi^\mu \partial_\mu = \partial_\phi\,.
\end{align}
From these, we may construct the currents
\begin{align}
    j_{(t)}^\mu = T^\mu{}_\nu t^\nu\,, \quad j_{(\phi)}^\mu = T^\mu{}_\nu \phi^\nu\,,
\end{align}
which are both conserved,
\begin{equation}
    \label{eq:killing_current_diff_law}
    \nabla_\mu j_{(i)}^\mu = 0\, ,
\end{equation}
a result of the Killing equation and energy-momentum conservation equation $\nabla_\mu T^{\mu\nu} = 0$. 
Retaining the use of the generic label $i\in \{t, \phi\}$, we may convert 
Eq.~\eqref{eq:killing_current_diff_law}
into an integral conservation law
by integrating it over a spacetime region $\mathcal{V}$, 
\begin{align}
    \nonumber
    0 = {}& \int_\mathcal{V} \dif^4x \sqrt{-g} \nabla_\mu j_{(i)}^\mu\\
    = {}& \oint_{\partial \mathcal{V}} \dif \Sigma_\mu j^\mu_{(i)}\, ,
    \label{eq:j_cons_law_general}
\end{align}
where we applied the divergence theorem in the second line.
To make this law useful, we must divide up the boundary $\partial\mathcal{V}$. We introduce four 3-surfaces: the horizon surface $\mathcal{H} = \{\mathcal{M}: x^1 = r_+ \}$, a timelike surface $\mathcal{B}_r = \{\mathcal{M}: x^1 = r>r_+ \}$, and two spacelike surfaces at $t=t_1, t_2$, {$\Sigma_{j} = \{\mathcal{M}: x^0 = t_{j} \}$}.
We define $\mathcal{V}$ as the region enclosed by these surfaces, shown in blue in Figure~\ref{fig:accretion_surfaces_sketch}.
We define the truncated surfaces $\{\tilde{\Sigma}_{(j)}, \tilde{\mathcal{H}}, \tilde{\mathcal{B}}_r\} = \{\Sigma_{(j)}, \mathcal{H}, \mathcal{B}_r\} \cap \partial \mathcal{V}$, which together form the boundary of $\mathcal{V}$, as shown in Figure~\ref{fig:accretion_surfaces_sketch}. We may thus split up Eq. \eqref{eq:j_cons_law_general} into
\begin{align}
    \nonumber
    0 = {}& \int_{\tilde{\mathcal{H}}} \dif \Sigma_\mu j_{(i)}^\mu - \int_{\tilde{\Sigma}_1} \dif \Sigma_\mu j_{(i)}^\mu
    \\ &
    + \int_{\tilde{\Sigma}_2} \dif \Sigma_\mu j_{(i)}^\mu - \int_{\tilde{\mathcal{B}}_r} \dif \Sigma_\mu j_{(i)}^\mu\, ,
    \label{eq:killing_charge_cons_law}
\end{align}
where $\dif \Sigma_\mu$ is the directed area element on each surface. We pick $\dif \Sigma_\mu$ to point ``forward" for the spacelike hypersurfaces and ``outward" for the timelike hypersurface. The locations of
these surfaces within the $(t,r)$ plane are shown in Figure~\ref{fig:accretion_surfaces_sketch}. 

The spacelike integrals in Eq.~\eqref{eq:killing_charge_cons_law} correspond to charges enclosed within their respective spatial regions, while the $\tilde{\mathcal{H}}$ and $\tilde{\mathcal{B}}_r$ integrals represent fluxes of these charges across their respective surfaces. In other words, the charge enclosed in our spatial region at coordinate time $t=t_2$ (third integral) is equal to the charge at $t=t_1$ (second integral) minus the charge transferred across and $\mathcal{B}_r$ surface (fourth integral) and across the \bh{} horizon (first integral).
In this work, both the background spacetime and the field's energy-momentum tensor are stationary, meaning the second and third integrals, over the $\tilde{\Sigma_j}$ surfaces, precisely cancel each other. 
We are thus left with the charge transfer law
\begin{equation}
    \label{eq:charge_transfer_law}
    \int_{\tilde{\mathcal{H}}} \dif \Sigma_\mu j_{(i)}^\mu = \int_{\tilde{\mathcal{B}}_r} \dif \Sigma_\mu j_{(i)}^\mu\, .
\end{equation}
This tells us that the charge transferred into the \bh{} in a given interval of coordinate time $t$ 
is equal to the charge that enters $\mathcal{V}$ across the outer timelike surface $\mathcal{B}_r$ in that same interval. We may thus define a rate associated with this transfer, or a ``flux", given by
\begin{align}
    \label{eq:killing_flux_formula}
    \mathcal{F}_{(i)} = {}& \frac{1}{\Delta t} \int_{\tilde{\mathcal{H}}} \dif \Sigma_\mu j_{(i)}^\mu = \frac{1}{\Delta t} \int_{\tilde{\mathcal{B}}_r} \dif \Sigma_\mu j_{(i)}^\mu\\\nonumber
    = {}& \frac{1}{\Delta t} \int_{\tilde{\mathcal{B}}_r} \dif y^3 \sqrt{|\gamma|}n_\mu j_{(i)}^\mu\, ,
\end{align}
where $y^a$, $\gamma_{ab}$, and $n_\mu$ are the coordinates, induced metric, and normal vector for the timelike hypersurface $\mathcal{B}_r$, and $\Delta t = t_2-t_1$. For the timelike surface $\mathcal{B}_r$, we have $\gamma = \det(\gamma_{ab}) = -\sin^2\theta \Delta \Sigma$ and $n_\mu \dif x^\mu = \sqrt{\frac{\Sigma}{\Delta}}\, \dif r$. Thus,
\begin{align}
    \nonumber
    \mathcal{F}_{(i)} = {}& \frac{1}{\Delta t} \int_{t_1}^{t_2}\dif t \oint_{S^2} \dif\Omega\, \Sigma\, T^r{}_{i} |_{r}\\
    = {}& \oint_{S^2} \dif\Omega \left(r^2 + a^2 \cos^2\theta\right) \left \langle T^r{}_{i} \right \rangle |_{r}\, ,
    \label{eq:flux_general_formula}
\end{align}
where $\langle \cdot \rangle$ denotes time-averaging over the interval $\Delta t$ and any $r$ outside the horizon can be chosen, as shown by Eq.~\eqref{eq:charge_transfer_law}.

The $\tilde{\Sigma}_j$ integrals in Eq. \eqref{eq:killing_charge_cons_law} (second and third integrals) represent amounts of Killing charge contained in the spatial regions $\tilde{\Sigma}_j$. Standard Minkowski field theory intuition leads us to define a Killing mass and Killing angular momentum associated with them, viz., 
\begin{subequations}
\label{eq:killing_charges_M_J}
\begin{align}
    \label{eq:mass_killing_charge}
    M_\text{Kil.} ={}& \int_{\tilde{\Sigma}_i} \dif \Sigma_\mu j_{(t)}^\mu\\
    \label{eq:angular_momentum_killing_charge}
    J_\text{Kil.} ={}& -\int_{\tilde{\Sigma}_i} \dif \Sigma_\mu j_{(\phi)}^\mu\, ,
\end{align}
\end{subequations}
where we pick the sign in the latter to recover the usual right-hand rule. 
All Killing charge lost from the region $\mathcal{V}$ across the horizon surface $\mathcal{H}$ must be added to the total Killing charge contained in the horizon. Thus, transfers of a given amount of Killing mass or angular momentum, Eqs.~\eqref{eq:killing_charges_M_J}, into the horizon increase the \bh{} mass or angular momentum by an equivalent amount.
This can be seen, e.g., by identifying the Killing charge contributions in the Komar mass and angular momentum formulae. 
Thus, the rates of change of the \bh{} mass and angular momentum are given by the horizon Killing fluxes of these quantities. Matching the sign of the Killing flux formula, Eq.~\eqref{eq:flux_general_formula}, for $\dot{M}$ and for $\dot{J}$ to the signs in Eqs.~\eqref{eq:killing_charges_M_J}, such that the conservation law in Eq.~\eqref{eq:j_cons_law_general} is reproduced, the \bh{} mass and angular momentum accretion rates are
\begin{subequations}
\begin{align}
    \label{eq:Mdot_general_formula}
    \dot{M} ={}& \mathcal{F}_{(t)} =  \oint_{S^2} \dif\Omega \left(r^2 + a^2 \cos^2\theta\right) \left \langle T^r{}_{t} \right \rangle |_{r}\\
    \label{eq:Jdot_general_formula}
    \dot{J} = {}& -\mathcal{F}_{(\phi)}\\\nonumber
    ={}& -\oint_{S^2} \dif\Omega \left(r^2 + a^2 \cos^2\theta\right) \left \langle T^r{}_{\phi} \right \rangle |_{r}\, .
\end{align}
\end{subequations}
These formulae are valid for any perturbation of the Kerr spacetime with a stationary energy-momentum tensor.

\section{Ratio of mass to angular momentum accretion rates}
\label{sec:Mdot_to_Jdot}

Calculations of the \bh{} mass and angular momentum accretion rates are greatly expedited if a determined relationship between them is known, allowing one to only calculate one quantity in full, 
then find the other using known formula.
In this section, we derive such a formula, relating $\dot{M}$ to $\dot{J}$ by a simple proportionality.
Modes of definite $(\omega, m)$ values form representations of the spacetime isometry group, characterized by the generators $\partial_t, \partial_\phi$. For a field populated in only one such mode, $A_\mu \sim e^{-i \omega t}e^{im \phi}$, its Lie derivatives along the generators are
\begin{align}
    \Lie_t A_\mu = -i \omega A_\mu\, , \quad \Lie_\phi A_\mu = i m A_\mu\, .
\end{align}
These modes thus carry definite $(t,\phi)$ Killing charges, Eqs. (\ref{eq:mass_killing_charge}-\ref{eq:angular_momentum_killing_charge}), with ratio $\omega / m$, and the accretion rates in Eqs. (\ref{eq:Mdot_general_formula}-\ref{eq:Jdot_general_formula}) thus vary by the same ratio~\cite{Bekenstein:1973mi}. This argument is completely general and applies to any type of test field in the Kerr spacetime. It is, however, rather abstract, and it is difficult to see immediately for a field of higher spin than a scalar. Here, we thus derive it explicitly for a Proca field in the Kerr spacetime, deriving a ratio between the field's mass and angular momentum accretion rates. 

The Killing flux through a spherical shell is given by Eq. \eqref{eq:flux_general_formula}, where the energy-momentum tensor for a Proca field is Eq. \eqref{eq:ProcaTmunu}. For an isometry generator $\xi^\mu_{(i)}$ with $i \in \{t, \phi\}$ and an eigenvalue $\lambda_{(i)} \in \{-i \omega, i m\}$, the integrand of the flux formula, Eq.~\eqref{eq:flux_general_formula}, reads
\begin{equation}
    \label{eq:flux_integrand_exp}
    n^\mu T_{\mu\nu} \xi_{(i)}^\nu = n^\mu \left(F_{(\mu}{}^{\rho} F^{\ast}_{\nu)\rho} \xi^\nu_{(i)}
+ \mu^2 A^{\ast}_{(\mu} A^{}_{\nu)} \xi^\nu_{(i)} \right)\, ,
\end{equation}
where $n^\mu \partial_\mu = \sqrt{\frac{\Delta}{\Sigma}} \partial_r$ is the spacelike normal from the previous section, and the last term in $T_{\mu\nu}$ vanishes because $n^\mu$ and $\xi^\mu_{(i)}$ are orthogonal, true for both Killing vectors. We begin with the first term. Without the symmetrization, we see
\begin{align}
    \label{eq:FFxi_1}
    F^\ast_\mu{}^\rho F_{\nu\rho} \xi^\nu_{(i)} ={}& F^\ast_\mu{}^\rho \left(\Lie_{\xi_{(i)}} A_\rho - \xi^\nu_{(i)} \nabla_\rho A_\nu \right)\\\nonumber
    ={}& \lambda_{(i)} F^\ast_\mu{}^\rho A_\rho - \left(\nabla_\mu A^{\ast \rho} - \nabla^\rho A^\ast_\mu \right) \xi^\nu_{(i)} \nabla_\rho A_\nu\, .
\end{align}
This term splits into a part proportional only to $\lambda_{(i)}$ and a part with a more complicated relationship on $\xi_{(i)}^\mu$. Since the spherical integral in Eq. \eqref{eq:flux_general_formula} may be taken at any $r$ without loss of generality, we may take advantage of the spacetime's asymptotic flatness by taking $r \to \infty$, simplifying Eq.~\eqref{eq:FFxi_1} to
\begin{align}
    F^\ast_\mu{}^\rho F_{\nu\rho} \xi^\nu_{(i)} ={}& 
    \lambda_{(i)} F^\ast_\mu{}^\rho A_\rho - \left(\partial_\mu A^{\ast \rho} - \partial^\rho A^\ast_\mu \right) \xi^\nu_{(i)} \partial_\rho A_\nu
\nonumber\\ ={}&
    \lambda_{(i)} F^\ast_\mu{}^\rho A_\rho + \partial_\rho\left(A^\ast_\mu \partial^\rho A_{(i)} - A_{(i)} \partial_\mu A^{\ast \rho} \right)
\nonumber\\  & 
    - A^\ast_\mu \partial_\rho\partial^\rho A_{(i)}
\,,
\end{align}
where $A_{(i)} = \xi^\mu_{(i)} A_\mu$, and we used the Lorenz condition in Eq.~\eqref{eq:Lorenzgauge}. Since $F^\ast_\nu{}^\rho F_{\mu\rho} \xi^\nu_{(i)} = \left(F^\ast_\mu{}^\rho F_{\nu\rho} \xi^\nu_{(i)}\right)^*$, 
reinstating the symmetrization amounts to simply taking the real part of this expression. We have thus reduced the first term in Eq.~\eqref{eq:flux_integrand_exp} to one well-behaved term, one total derivative term, and one term which must be eliminated. 

Again setting aside the symmetrization, the second term in Eq. \eqref{eq:flux_integrand_exp} can be rewritten using the Proca equation \eqref{eq:ProcaEoM} to give
\begin{equation}
    \mu^2 A^\ast_\mu A_\nu \xi_{(i)}^\nu = -A^\ast_\mu \nabla_\rho F_\nu{}^\rho \xi_{(i)}^\nu\, .
\end{equation}
Using our choice $r \to \infty$ and the Lorenz condition, we find
\begin{equation}
    \mu^2 A^\ast_\mu A_\nu \xi_{(i)}^\nu = A_\mu^\ast \partial_\rho \partial^\rho A_{(i)}\, .
\end{equation}
Symmetrizing again amounts to taking the real part of this expression. Thus, using the $r\to \infty$ limit,
for which
$n^\mu \partial_\mu = \sqrt{\frac{\Delta}{\Sigma}} \partial_r \to \partial_r$, the full integrand in Eq.~\eqref{eq:flux_integrand_exp} reads
\begin{align}
    n^\mu T_{\mu\nu} \xi_{(i)}^\nu ={}& \re \bigl( \lambda_{(i)}F^\ast_r{}^\mu A_\mu\bigr) + \partial_\mu \mathcal{T}_{(i)}^\mu\, 
\end{align}
where
\begin{equation}
    \mathcal{T}_{(i)}^\mu \coloneq \re \bigl(A^\ast_r \partial^\mu A_{(i)} - A_{(i)} \partial_r A^{\ast \mu}\bigr)\, .
\end{equation}
Using Eq. \eqref{eq:flux_general_formula}, the full Killing flux is
\begin{align}
    \mathcal{F}_{(i)} ={}& 
    \oint \dif \Omega\, r^2 \re \bigl(\lambda_{(i)}\left\langle F_r^\ast{}^\rho A_\rho \right\rangle\bigr)\\\nonumber
    &+ \oint \dif \Omega\, r^2 \left\langle \partial_t \mathcal{T}_{(i)}^t \right\rangle \, ,
\end{align}
where the spherical integral eliminates the angular part of $\partial_\mu \mathcal{T}_{(i)}^\mu$, and the radial part vanishes because $\mathcal{T}^r_{(i)} = 0$. 
If the frequency is real, $\omega \in \mathbb{R}$, then the sole time-dependent factor in $\mathcal{T}^\mu_{(i)}$, $e^{i(\omega^\ast - \omega) t}$, extracted from $A_\mu \sim e^{-i \omega t}$, reduces to unity, and $\partial_t \mathcal{T}^t_{(i)} = 0$. 
In the general case of complex frequencies, $\omega \in \mathbb{C}$,
a time-dependent factor $e^{2\omega_I t}$ remains, meaning the term does not vanish identically nor after the time average. However, if $|\omega_I| \ll |\omega_R|$, as for a long-lived bound state, then the term is small relative to the former integral and can be neglected. 

Putting these together, for sufficiently small $\omega_I$ we see
\begin{align}
    \frac{\dot{J}}{\dot{M}} = \frac{-\mathcal{F}_{(\phi)}}{\mathcal{F}_{(t)}} = \frac{- m \oint \dif \Omega\, r^2 \left\langle \re \bigl(F_r^\ast{}^\rho A_\rho \right\rangle}{-\omega_R \oint \dif \Omega\, r^2 \left\langle \re \bigl(F_r^\ast{}^\rho A_\rho \right\rangle}\, .
\end{align}
Thus,
\begin{equation}
    \label{eq:Jdot_to_Mdot}
    \dot{J} = \frac{m}{\omega_R} \dot{M}\, ,
\end{equation}
which is our final result. 
We note that in this work, we choose
$\omega = \mu \in \mathbb{R}$, for which this relation is exactly true.

\bibliographystyle{apsrev4-2}
\bibliography{Refs_ProcaHair.bib}

\end{document}